\documentclass[]{aa} 

\usepackage{epsfig}

\renewcommand{\d}{{\rm d}}
\newcommand{\pl}{\partial}
 
\newcommand{\beq}{\begin{equation}} 
\newcommand{\eeq}{\end{equation}} 
\newcommand{\beqa}{\begin{eqnarray}} 
\newcommand{\eeqa}{\end{eqnarray}} 
\newcommand{\bea}{\begin{array}} 
\newcommand{\ea}{\end{array}} 
\newcommand{\cG}{{\cal G}} 
\newcommand{\rhob}{\overline{\rho}} 
\newcommand{\lag}{\langle} 
\newcommand{\rag}{\rangle}
\newcommand{\bx}{{\bf x}}

\newcommand{\bv}{{\bf v}}

\newcommand{\varphib}{\overline{\varphi}}

\begin{document} 
 
\title{Thermodynamics and dynamics of a 1-D gravitational system}    
\author{P. Valageas}   
\institute{Service de Physique Th\'eorique, CEN Saclay, 91191 Gif-sur-Yvette, 
France}  
\date{Received 4 November 2005 / Accepted 11 January 2006} 

\abstract
{}
{The dynamics of large-scale structure formation in the universe by 
gravitational instability still presents many open issues and is mostly 
studied through numerical simulations. This motivates the study of simpler 
models which can be investigated by analytical means in order to understand 
the main processes at work. Thus, we describe here a one-dimensional 
self-gravitating system derived from the cosmological context, which leads to
an effective external potential in addition to the standard gravitational 
self-interaction. As a first step we consider small times so that the 
expansion can be neglected. Then we present a thermodynamical analysis of this 
system as well as the stability properties of the associated hydrodynamical 
and collisionless systems.}
{We consider the mean field limit (i.e. continuum limit) to perform an analytical 
study.}
{We find a second-order phase transition at $T_{c1}$ from an homogeneous 
equilibrium at high temperature to a clustered phase (with a density 
peak at one of the boundaries of the system) at low temperature. There also 
exists an infinite series of unstable equilibria which appear at lower
temperatures $T_{cn}$, reflecting the scale-free nature of the gravitational
interaction and the usual Jeans instability. We find that, as for the similar
HMF (Hamiltonian mean field) model, all three micro-canonical, canonical and
grand-canonical ensembles agree with each other, as well as with the stability
properties associated with a hydrodynamical approach. On the other hand, the
collisionless dynamics governed by the Vlasov equation yields the same results
except that at low $T$ the equilibrium associated with two density peaks (one
at each boundary) becomes stable.}
{}

 
\maketitle

\section{Introduction} 
\label{Introduction}

In standard cosmological scenarios the large scale structures we observe in
the present universe (such as clusters, filaments or voids) have formed through
the amplification by gravitational instability of small primordial perturbations,
see Peebles (1980). Moreover, the amplitude of these density fluctuations 
increases
at small scales as in the CDM model (Peebles 1982) which leads to a hierarchical
process where smaller scales become non-linear first. Then, at large scales or
at early times one can use a perturbative approach to study the evolution of
initial fluctuations (Fry 1984; Goroff et al. 1986; Bernardeau 1992; 
Valageas 2001). 
Next, the weakly non-linear regime may be investigated through the Zeldovich
approximation (Shandarin \& Zeldovich 1989) or the adhesion model 
(Gurbatov et al. 1989).
However, the highly non-linear regime (which corresponds to collapsed structures 
such as clusters of galaxies) has mostly remained out of reach of systematic
approaches. Thus, one may use the Press-Schechter approximation 
(Press \& Schechter 1974)
or the excursion-set formalism of Bond et al. (1991) to obtain the statistics of 
just-virialized halos or the saddle-point approach of Valageas (2002) for rare
voids. These approximations focus on specific objects which may be ``recognized''
from individual features in the linear density field itself and do not follow
the system as a whole through its non-linear evolution. A systematic method
to do so was recently developed in Valageas (2004) but its application to the
collisionless dynamics has not been performed yet for the highly non-linear 
regime. Therefore, the non-linear regime of cosmological structure formation
is mostly studied through numerical simulations.

In order to simplify this difficult problem one can investigate one-dimensional
(1-D) systems which
are easier to study both through numerical and analytical approaches. In fact,
1-D systems such as the system of parallel mass sheets (Camm 1950) have been 
studied for a long time to explore the evolution of isolated $N-$body systems 
which only interact through classical gravity. In particular, they have been used
to investigate relaxation processes and to test Lynden-Bell's prediction 
for the final state of violent relaxation (Lynden-Bell 1967). Such 1-D systems can
already exhibit complex behaviours. For instance, Luwel \& Severne (1985) found
that collisional effects are not sufficient to relax the stationary waterbag 
configuration towards thermodynamical equilibrium whereas Rouet \& Feix (1999) 
showed that holes in phase-space in the initial distribution function can 
persist over long times and prevent efficient relaxation.
On the other hand, the system can break up into smaller clusters as in 
Hohl \& Feix (1967) or develop a fractal structure as in Koyama \& Konishi (2001).

In this paper, we consider the 1-D gravitational system obtained by studying
1-D density fluctuations in a 3-D cosmological background. This model has
already been studied mostly through numerical simulation in 
Aurell \& Fanelli (2001), 
Aurell et al. (2001) and Fanelli \& Aurell (2002), both with and without 
cosmological expansion.
Here we restrict ourselves to time-scales much smaller than the Hubble time 
so that the expansion of the universe can be neglected and we focus on a mean 
field analysis (i.e. a continuum limit) which is relevant in the cosmological 
context. We study the thermodynamics and stability of this system, similarly to
the theoretical analysis of Chavanis et al. (2005) performed for the HMF
model (defined by a cosine interaction) which shows a similar behaviour. We 
introduce this model in sect.~\ref{cosmological-gravitational-system}, where 
the effect of the cosmological background is seen to reduce to an effective 
external potential $V$ which balances the gravitational self-interaction $\Phi$ 
so that the homogeneous state is an equilibrium solution. Next, we perform a 
thermodynamical analysis of this system in sect.~\ref{Thermodynamical-analysis}, 
for the micro-canonical, canonical and grand-canonical ensembles. We study both 
the possible equilibrium states and their stability properties. Then, in 
sect.~\ref{Hydrodynamical-model} we investigate the stability of such equilibria
within an hydrodynamical framework whereas we consider the collisionless
case as described by the Vlasov dynamics in sect.~\ref{Vlasov}. Finally, we
present our conclusions in sect.~\ref{Conclusion}.

\section{A 1-D cosmological gravitational system}
\label{cosmological-gravitational-system}

The formation of large-scale structures in the universe through gravitational
instability is described by the Vlasov-Poisson system which applies to
a collisionless fluid in the continuum limit where the mass $m$ of the 
particles goes to zero. Thus, the distribution function (phase-space density)
$f(\bx,\bv,t)$, where we note $\bx$ the comoving coordinate of a particle and 
$\bv$ its peculiar velocity defined by $\bv = a^2 \dot{\bx}$ (and $a(t)$ is
the cosmological scale-factor), obeys the Vlasov equation (see Peebles 1980):
\beq
\frac{\pl f}{\pl t} + \frac{\bv}{a^2} . \frac{\pl f}{\pl \bx} -
\frac{\pl \phi}{\pl \bx} . \frac{\pl f}{\pl \bv} = 0 ,
\label{Vlasov1}
\eeq
while the gravitational potential $\phi$ satisfies the Poisson equation:
\beq
\Delta \phi = \frac{4\pi \cG}{a} (\rho-\rhob) \;\;\; \mbox{with} \;\;\;  
\rho(\bx,t) = \int f(\bx,\bv,t) \; \d\bv .
\label{Poisson1}
\eeq
Here we introduced the comoving density $\rho(\bx,t)$ and its average $\rhob$
(which is independent of time), and we note $\cG$ Newton's constant. The fact 
that the Poisson equation involves the density perturbation $(\rho-\rhob)$ 
rather than the total density $\rho$ comes from the change from physical 
coordinates to comoving coordinates. In particular, for an homogeneous
universe which merely follows the Hubble flow we have $\phi=0$ and
$f(\bx,\bv,t)=\rhob \delta_D(\bv)$ where $\delta_D$ is Dirac's distribution.
If we consider planar perturbations along the axis $x$: 
$f(\bx,\bv,t)= f(x,v,t) \delta_D(\bv_{\perp})$, we obtain a one-dimensional 
(1-D) problem:
\beq
\frac{\pl f}{\pl t} + \frac{v}{a^2} . \frac{\pl f}{\pl x} -
\frac{\pl \phi}{\pl x} . \frac{\pl f}{\pl v} = 0 , \;\;\;
\frac{\pl^2\phi}{\pl x^2} = \frac{4\pi \cG}{a} (\rho-\rhob) .
\label{Vlasov2}
\eeq
Let us moreover restrict ourselves to periodic systems of period $2L$ which
are symmetric with respect to the origin $x=0$. Then, by symmetry the force
$F=-\pl\phi/\pl x$ vanishes at the origin, $F(x=0)=0$, and the gravitational 
potential can be integrated as: 
\beq
F(x)= -\frac{\pl\phi}{\pl x} = \frac{2\pi \cG}{a} \int_0^L \d x' 
(\rho-\rhob)(x') \; {\rm sign}(x'-x) ,
\label{F1}
\eeq
and:
\beq
\phi(x)= \frac{2\pi \cG}{a} \int_0^L \d x' (\rho-\rhob)(x') \; |x-x'| ,
\label{phi1}
\eeq
where we restrict ourselves to $0\leq x\leq L$. If we consider time-scales
which are much smaller than the Hubble time $H^{-1}=a/\dot{a}$ we can
neglect the time dependence of the scale-factor $a(t)$ and taking $a=1$ and
$g=2\pi\cG$ we can write the full potential $\phi$ as:
\beq
\phi=\Phi+V \;\; \mbox{with} \; \left\{ \bea{l} \Phi= g \int_0^L\d x'\rho(x') 
\; |x-x'| \\ 
V(x)= -g \rhob \left[ \left(x-\frac{L}{2}\right)^2 + \frac{L^2}{4} \right]   
\ea \right.
\label{phi2}
\eeq
where we split $\phi$ into a standard gravitational ``interaction'' term 
$\Phi$ and an ``external'' potential $V$. Thus, the associated discrete system
of $N$ particles of finite mass $m$ is described by the Hamiltonian 
${\cal H_N}$:
\beqa
{\cal H_N} & = & \sum_{i=1}^N \frac{1}{2} m v_i^2 + g m^2 \sum_{i>j} |x_i-x_j|
\nonumber \\ && - g \rhob m \sum_{i=1}^N 
\left[ \left(x_i-\frac{L}{2}\right)^2 + \frac{L^2}{4} \right] .
\label{HN}
\eeqa
Alternatively, the mean field approximation to the $N-$body system (\ref{HN})
is given by the potential (\ref{phi2}) and the Vlasov-Poisson equations.
The important feature of this one-dimensional static cosmology (OSC) model
which is due to its cosmological setting is the appearance of the ``background''
potential $V(x)$ which ensures that the homogeneous configuration $\rho=\rhob$
is a solution of the equation of motion. This is the counterpart of the
Hubble flow. The OSC model defined by (\ref{HN}) or (\ref{phi2})
was studied numerically in Aurell et al. (2001) and Aurell \& Fanelli (2001)
(where it is called a ``frictionless model'') for random (Brownian motion)
and non-random (sinusoidal) initial conditions. 
In this article, we shall consider the thermodynamics of the continuous
system (\ref{phi2}), which can also be seen as the mean field approximation 
to (\ref{HN}). This is a system of finite size $L$ with $0\leq x\leq L$.
However, it can be useful to remember that it can also be understood as a 
periodic system of size $2L$ which is symmetric with respect to the origin 
$x=0$.

Here, we should add a few words about the limitations of this OSC model
with regard to cosmology. The real universe is obviously three-dimensional and
some important features such as the support of self-gravitating halos by rotation
(as for spiral galaxies) cannot be studied through 1-D models. Thus, although 
the early stages of non-linear evolution are governed by 1-D collapse 
(formation of pancakes, Shandarin \& Zeldovich 1989) the highly
non-linear regime which is of interest to us is truly three-dimensional.
However, we can expect that key processes associated with the scale-free nature
of the gravitational interaction are common to 1-D, 2-D and 3-D systems. For
instance Munshi et al. (1999) find that 2-D and 3-D cosmological simulations 
exhibit similar behaviours in the highly non-linear regime. In particular,
1-D systems should help to check whether the density profiles of virialized
halos show a universal shape, for a wide class of initial conditions, and whether
the power-spectrum of density fluctuations saturates into the non-linear regime
or exhibits a power-law growth which could depend on initial conditions.
A study of 1-D systems such as the OSC model is particularly timely as modern
supercomputers are beginning to allow 1-D cosmological simulations in phase-space
instead of the usual $N-$body techniques (Alard \& Colombi 2005).

On the other hand, the use of a finite size domain $0\leq x\leq L$ (which is
equivalent to periodic boundary conditions of period $2L$ with symmetry with
respect to $x=0$) is customary to studies of gravitational systems 
(e.g. Padmanabhan 1990). Within the cosmological context, characterized by a
hierarchical scenario where smaller scales turn non-linear first, the size $L$ 
has actually a physical meaning. It should be interpreted as the wavelength 
which turns non-linear at the time of interest. Indeed, larger scales which are 
still linear follow the Hubble flow (up to small perturbations) while smaller 
scales have relaxed within virialized halos. Thus, to this scale is also 
associated a characteristic time $t_H$ given by the age of the universe when
$L$ enters the non-linear regime. Today this yields $t_H \sim 10^{10}$ yrs
and $L \sim 5$ Mpc. Then, the expansion of the universe can be neglected if
we restrict ourselves to length and time scales much smaller than $L$ and $t_H$.
Of course, this may also prevent the system to reach thermodynamical equilibrium
which can require very long timescales (increasing algebraically with 
the number $N$ of particles, Chavanis 2006).

However, this article should only be seen as a first step in the study of
the OSC system. We think that this model is also interesting 
in its own right as it should exhibit the features associated with a long-range
scale-free interaction, which is an active domain of research. Besides, the 
background $-\rhob$ in Poisson 
eq.(\ref{Poisson1}) which arises from the cosmological context also leads to
interesting transition phenomena as described in 
sect.~\ref{Thermodynamical-analysis} below.

\section{Thermodynamical analysis}
\label{Thermodynamical-analysis}

\subsection{Statistical equilibrium}
\label{Statistical-equilibrium}

\subsubsection{Micro-canonical ensemble}
\label{Micro-canonical-ensemble1}

\paragraph{Equilibrium profiles:}
\label{Equilibrium-profiles:}
$\hfill$

In the micro-canonical ensemble the statistical equilibrium is obtained
by maximizing the Boltzmann entropy $S[f]$ at fixed energy $E$ and mass $M$,
given by (Padmanabhan 1990, Chavanis et al. 2005):
\beq
S = - \int \d x \d v \; f \ln f , \; \;\;
E = \int \d x \d v \; f \left( \frac{v^2}{2} + \frac{\Phi}{2} + V \right)
\label{S1}
\eeq
and: 
\beq
M = \rhob L = \int\d x \d v f .
\label{M}
\eeq
Using Lagrange multipliers this yields the usual Maxwell-Boltzmann 
distribution:
\beq
f(x,v) \propto e^{-\beta[v^2/2+\phi(x)]} \;\;\; \mbox{and} \;\;\; 
\rho(x) \propto e^{-\beta \phi(x)} ,
\label{Maxwell}
\eeq
where we introduced the inverse temperature $\beta=1/T$. Let us define the 
offset $\varphib$ such that:
\beq
\phi=\varphi+\varphib \;\; \mbox{with} \;\; \rho = \rhob e^{-\beta \varphi}
\;\; \mbox{and} \;\; \frac{\d^2\varphi}{\d x^2} = 2 g (\rho-\rhob) .
\label{varphib}
\eeq
Therefore, the thermodynamical equilibrium is set by the equations:
\beq
\frac{\d^2\varphi}{\d x^2} = 2 g \rhob \left( e^{-\beta\varphi}-1 \right) 
\;\;\; \mbox{and} \;\;\; \varphi'(0)=\varphi'(L)=0 ,
\label{LE1}
\eeq
where the boundary conditions are obtained from eq.(\ref{F1}).
Without the constant term $-1$ in its r.h.s. eq.(\ref{LE1}) reduces to the 
usual Lane-Emden equation which arises for isolated self-gravitating isothermal
gaseous spheres (e.g., Chandrasekhar 1942; Binney \& Tremaine 1987).
It is convenient to introduce the Jeans length $L_J$ (Binney \& Tremaine 1987) 
and the ratio $\zeta_L$:
\beq
L_J = \left( 2g\rhob\beta \right)^{-1/2} , \;\;\; 
\zeta_L= \frac{L}{L_J} = \sqrt{2g\rhob\beta} L  ,
\label{LJzetaL}
\eeq
in order to define the dimensionless quantities:
\beq
\zeta = \frac{x}{L} \zeta_L , \;\;  \psi= \beta \varphi \;\; \mbox{and} 
\;\; \eta = \frac{\rho}{\rhob} \; ,\;\; \mbox{whence} \;\;
\eta = e^{-\psi} .
\label{psieta} 
\eeq
Then, the modified Lane-Emden eq.(\ref{LE1}) reads:
\beq
\frac{\d^2\psi}{\d\zeta^2} = e^{-\psi}-1 \;\;\; \mbox{and} \;\;\;
\psi'(0)=\psi'(\zeta_L)=0 .
\label{LE2}
\eeq
This is the equation of motion of a particle of mass $m=1$ and coordinate
$\psi$ as a function of time $\zeta$ within a potential 
${\cal V}(\psi)= e^{-\psi}+\psi-1$, with zero velocity at initial ($\zeta=0$)
and final ($\zeta_L$) times. The case where the particle is at rest at the
bottom $\psi=0$ of the potential ${\cal V}(\psi)$ is always a solution
of eq.(\ref{LE2}). This corresponds to the uniform density $\rho=\rhob$.
As the temperature $T$ decreases, while $\beta$ and $\zeta_L$ grow, new 
solutions appear as the particle now has time to perform one or several
oscillations in the potential well ${\cal V}(\psi)$. This yields a series
of critical temperatures $T_{cn}$, or $\beta_{cn}$, associated with the
appearance of a solution with $n$ half-oscillations. They can be
obtained by expanding the potential ${\cal V}(\psi)$ around $\psi=0$ up
to quadratic order. This yields:
\beq
{\cal V}(\psi) \simeq \frac{\psi^2}{2} , \;\; \mbox{whence} \;\;  
\psi''+\psi=0 \;\; \mbox{and} \;\; \psi = \psi_0 \cos\zeta ,
\label{psi0}
\eeq
where we used $\psi'(0)=0$. Next, the second constraint $\psi'(\zeta_L)=0$
implies:
\beq
\zeta_L= n \pi \;\;\; \mbox{whence} \;\;\; 
\beta_{cn} = \frac{n^2\pi^2}{2g\rhob L^2} ,  \;\;\; 
T_{cn} = \frac{2g\rhob L^2}{n^2\pi^2}  .
\label{betacn}
\eeq
If $n$ is odd this corresponds to $n$ half-oscillations starting at 
$\psi_0$ and ending at $-\psi_0$ whereas if $n$ is even it corresponds to
$n/2$ full oscillations starting and ending at $\psi_0$. As the temperature
decreases below $T_{cn}$ the oscillations get wider and take ``more time''. 
In the following we shall label by ``$n$'' the solution with $n$ 
half-oscillations which starts at $\psi(0)=\psi_-<0$ and by $-n$ the solution 
with $n$ half-oscillations which starts at $\psi(0)=\psi_+>0$.
From the last equation in (\ref{Maxwell}) or (\ref{psieta}) we see for instance
that the mode $n=2$ shows two density peaks at $x=0$ and $x=L$ while the
mode $n=-2$ shows only one density peak at $x=L/2$ (but two underdense minima
at $x=0$ and $x=L$). We shall call the homogeneous equilibrium solution 
$\rho=\rhob$ as the mode $n=0$.
On the other hand, since the differential equation in
(\ref{LE2}) does not depend on $\beta$ or $n$ (which only appear in the
boundary conditions) all modes $\pm n$ can be written in terms of the mode 
$n=1$ at a different temperature $T_1$ given by $\zeta_{L1}=\zeta_{L}/n$ 
or $\beta_1=\beta/n^2$. Thus, we have:
\beqa
\psi_n(\zeta;\beta) & = & \psi_1\left(\zeta;\frac{\beta}{n^2}\right) 
\label{psin} \\
\psi_{-n}(\zeta;\beta) & = & 
\psi_1\left(\zeta-\frac{\zeta_L}{n};\frac{\beta}{n^2}\right)
\label{psi-n}
\eeqa
and similar relations for the overdensity $\eta$.
Here we considered that all solutions are extended over the full real axis
to yield solutions which are periodic of period $2L$ and are symmetric with 
respect to $x=0$, in agreement with 
sect.~\ref{cosmological-gravitational-system}. This merely amounts to follow 
the motion of the particle within the potential well ${\cal V}(\psi)$ over 
all times. The periodicity and symmetry of $\psi(\zeta)$ implies that we can 
expand the solution as:
\beq
\psi(\zeta) = \sum_{k=0}^{\infty} a_k 
\cos\left(k\pi\frac{\zeta}{\zeta_L}\right) .
\label{psik}
\eeq
This automatically satisfies the boundary conditions (\ref{LE2}). As seen in
eqs.(\ref{psin})-(\ref{psi-n}) it is sufficient to consider the case $n=1$.
Then, in order to study the behaviour close to the critical temperature
we can substitute the expansion (\ref{psik}) into the differential 
eq.(\ref{LE2}) up to the required order over $(\beta-\beta_{c1})$. We obtain
up to order $(\beta-\beta_{c1})$:
\beq
a_0= \frac{a_1^2}{4} , \;\;\; a_2= -\frac{a_1^2}{12} , \;\;\;
a_1 = - \sqrt{12 \frac{\beta-\beta_{c1}}{\beta_{c1}}} ,
\label{a1}
\eeq
while higher-order terms $a_k$ are of order $a_k \sim a_1^k$ for $k\geq 1$.
Thus we obtain a square-root singularity for the amplitude of the 
potential $\psi$ and for the density profile.

\begin{figure}
\begin{center}
\epsfxsize=8.5 cm \epsfysize=6.5 cm {\epsfbox{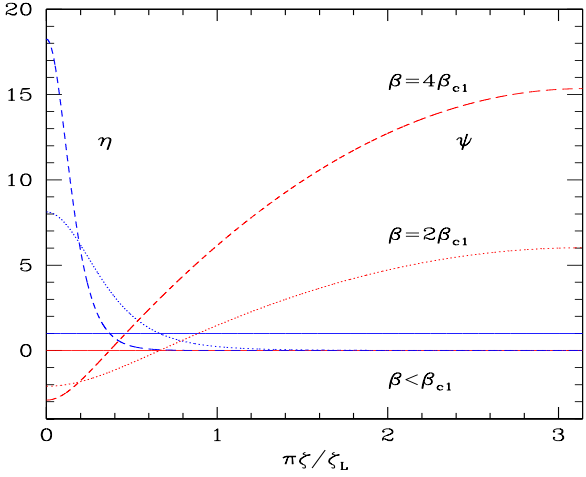}} 
\end{center}
\caption{The equilibrium profile $n=1$ obtained at $\beta=2\beta_{c1}$ 
(dotted lines) and $\beta=4\beta_{c1}$ (dashed lines). The solid lines
correspond to the uniform solution ($n=0$) with $\eta=1$ and $\psi=0$.
The normalized density profiles $\eta$ are the curves with a maximum at 
$\zeta=0$ while the normalized potentials $\psi$ have a maximum at 
$\zeta=\zeta_L$.}
\label{figprof}
\end{figure}

On the other hand, we can also obtain asymptotic expressions for the 
equilibrium density profile in the limit of low temperatures $T\rightarrow 0$.
Thus, close to $\zeta=0$ where $\psi\rightarrow-\infty$ we can approximate
${\cal V}(\psi)\simeq e^{-\psi}$ which yields from eq.(\ref{LE2})
\beq
\zeta \ll \frac{\ln\zeta_L}{\zeta_L} : \;\; \zeta \simeq \int_{\psi_-}^{\psi} 
\frac{\d\psi'}{\sqrt{2(e^{-\psi_-}-e^{-\psi'})}} ,
\label{zetapsi}
\eeq
which leads to:
\beq
\zeta \ll \frac{\ln\zeta_L}{\zeta_L} : \;\; 
\eta(\zeta) \simeq \frac{\zeta_L^2}{2\cosh^2(\zeta_L \zeta/2)} .
\label{etacosh}
\eeq
As expected, we recover the equilibrium profile obtained by Camm (1950) for
an isolated 1-D gravitational system. Indeed, close to the density peak
we can neglect $\rhob$ in Poisson equation (\ref{varphib}) in the low 
temperature limit (since $\rho(0) \sim 1/T$ from eq.(\ref{etacosh})).
Thus for arbitrary $n$ we obtain a series of density peaks which follow the
equilibrium profile of isolated 1-D gravitational systems separated by
low density regions since in the case of the mode $n=1$ we obtain:
\beq
\eta(\zeta_L)\sim e^{-\zeta_L^2/2}=e^{-\pi^2 T_{c1}/2T} ,
\label{etavoid}
\eeq
where we used the approximation ${\cal V}(\psi)\simeq \psi$ in the underdense 
region. We can note that the density in these underdense regions decreases 
very fast at low temperatures.

We display in fig.~\ref{figprof} the equilibrium profiles $n=0$ (homogeneous
solution) and $n=1$ (one density peak at $\zeta=0$) at temperatures
$T_{c1}/2$ and $T_{c1}/4$. We can already check at these temperatures that 
the density peak grows as $1/T$ while its width decreases as $T$ and that
the underdense region is almost void. Of course, the potential $\psi=-\ln\eta$
shows a much shallower shape.

\paragraph{Thermodynamical quantities:}
\label{Thermodynamical-quantities:}
$\hfill$

\begin{figure}
\begin{center}
\epsfxsize=8 cm \epsfysize=6 cm {\epsfbox{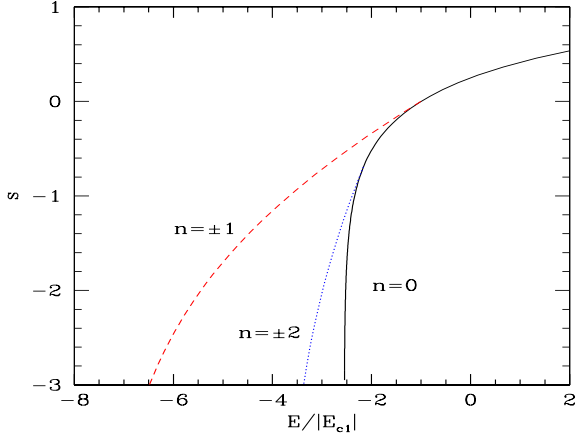}} 
\end{center}
\caption{The specific entropy $s=(S-S(T_{c1}))/M$ as a function of the 
energy $E$ for the modes $n=0,\pm 1$ and $\pm 2$. The lower energy bound
$E_{\rm min}(n)$ associated with mode $\pm n$, reached at $T=0$, increases for
higher $n>0$. The homogeneous solution $n=0$ has the highest bound 
$E_{\rm H}$. All thermodynamical quantities are identical for both equilibria
$n$ and $-n$.}
\label{figSE}
\end{figure}

\begin{figure}
\begin{center}
\epsfxsize=8 cm \epsfysize=6 cm {\epsfbox{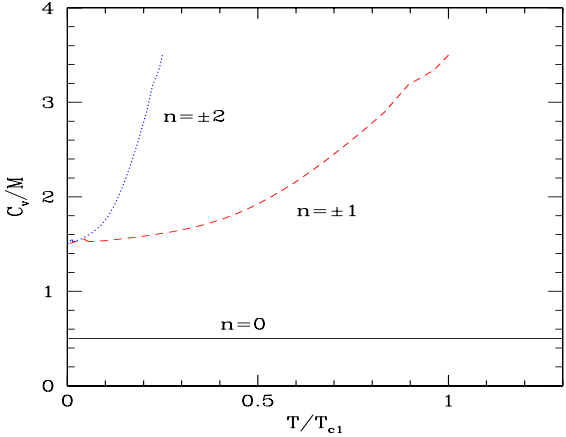}} 
\end{center}
\caption{The specific heat $C_v=\d E/\d T$ as a function of temperature for the
equilibria $n=0,\pm 1$ and $\pm 2$. It is discontinuous at the transitions 
$T_{cn}$ and corresponds to a second-order phase transition.}
\label{figCv}
\end{figure}

\begin{figure}
\begin{center}
\epsfxsize=8 cm \epsfysize=6 cm {\epsfbox{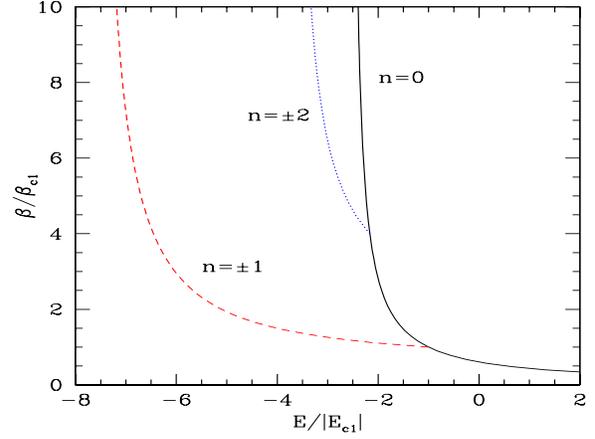}} 
\end{center}
\caption{The calorific curve $\beta(E)$ for the modes $n=0,\pm 1$ and $\pm 2$.
It shows a second-order phase transition at $T_{c1}$.}
\label{figbetaE}
\end{figure}

From eq.(\ref{Maxwell}) the distribution function at equilibrium 
is:
\beq
f(x,v) = \rho(x) \sqrt{\frac{\beta}{2\pi}} \; e^{-\beta v^2/2} ,
\label{Maxwell2}
\eeq
whence the entropy $S$ defined in eq.(\ref{S1}) is up to an additional 
constant:
\beq
S= \frac{M}{2} \ln T - \int \d x \; \rho \ln \frac{\rho}{\rhob} = 
\frac{M}{2} \ln T + \rhob \int \d x \; \eta \psi ,
\label{S2}
\eeq
where we used eqs.(\ref{psieta}). Therefore, the specific entropy $s$
normalized such that $s(T_{c1})=0$ is:
\beq
s= \frac{S-S(T_{c1})}{M} = \frac{1}{2}\ln\frac{T}{T_{c1}} 
+ \int_0^{\zeta_L} \frac{\d\zeta}{\zeta_L} \; \eta \psi .
\label{sT}
\eeq
On the other hand, the energy $E$ at thermodynamical equilibrium can be 
written from eq.(\ref{S1}) and eq.(\ref{Maxwell}) as:
\beq
E= \frac{MT}{2} + \int \d x \; \rho \frac{\phi+V}{2} \;\;\; \mbox{with} 
\;\;\; \rho=\rhob + \frac{1}{2g} \frac{\d^2\phi}{\d x^2} .
\label{E1}
\eeq
Integrating over the term $\rho V$, using the second eq.(\ref{E1}) and 
integration by parts, yields:
\beq
E= \frac{MT}{2} - \frac{gM^2L}{6} + \frac{M}{4} [\phi(0)+\phi(L)] 
+ \int \d x \; (\rho-\rhob) \frac{\phi}{2} .
\label{E2}
\eeq
From eq.(\ref{phi1}) we obtain $\phi(0)+\phi(L)=0$ therefore the energy of
modes $0$ and $\pm n$ reads:
\beqa
E_0 & = & \frac{MT}{2} - \frac{gM^2L}{6} 
\label{E0} \\
E_{\pm n} & = & \frac{MT}{2} - \frac{gM^2L}{6} + \frac{MT}{2}
\int_0^{\zeta_L} \frac{\d\zeta}{\zeta_L} \; (\eta-1)\psi .
\label{En}
\eeqa
In particular, the energy $E_{cn}$ associated with the critical temperature 
$T_{cn}$ of eq.(\ref{betacn}) is:
\beq
E_{cn} = - \frac{n^2\pi^2-6}{6n^2\pi^2} gM^2L \;\;\; \mbox{at} 
\;\;\; T_{cn} = \frac{2gML}{n^2\pi^2}  .
\label{Ec1}
\eeq
Here we can note that the energy $E$ of the OSC system is actually bounded 
from below. Indeed, as can be clearly seen from eq.(\ref{HN}) the minimum
value $E_{\rm min}$ for the energy is reached when all the mass is at rest
at $x=0$ or $x=L$. On the other hand the energy $E_{\rm H}$ associated with
the homogeneous static solution $\rho=\rhob$ and $v=0$, which would be the
counterpart of the original Hubble flow, can be obtained from eq.(\ref{E0})
with $T=0$. This yields:
\beq
E_{\rm min} = - \frac{gM^2L}{2} \;\;\; \mbox{and} \;\;\;
E_{\rm H} = - \frac{gM^2L}{6} .
\label{Emin}
\eeq
We can note that since $E_{\rm H}=E_0(T=0)$ all equilibrium solutions $\pm n$
can be reached with this energy, as may also be seen from fig.~\ref{figSE} where
we display the specific entropy $s$ as a function of the energy $E$.
As we can check in fig.~\ref{figSE} the mode $n=\pm 1$ reaches the energy
bound $E_{\rm min}$ at $T=0$ while other equilibrium states $\pm n$ are
increasingly close for larger $n$ to the homogeneous solution, with respect
to global quantities such as energy or entropy, since they are made of $n/2$
density peaks separated by a length $2\zeta_L/n$, and their minimal energy
$E_{\rm min}(n)$ gets closer to $E_{\rm H}$. Although fig.~\ref{figSE} only
shows the modes $0,\pm 1$ and $\pm 2$ we can see that at a fixed energy $E$ the
homogeneous state has the lowest entropy while modes $\pm n$ with $n\geq 1$
have a higher entropy as $n$ decreases. In particular, below $T_{c1}$
the mode $\pm 1$ with only one density peak at $x=0$ or $x=L$ is the highest
entropy equilibrium. Therefore, it is a stable equilibrium solution of the
system, as we shall also check in sect.~\ref{Micro-canonical-ensemble2},
whereas the homogeneous solution is unstable. Thus, as in the case of isolated
1-D systems (where only equilibrium $\pm 1$ exists and extends to all 
temperatures) and contrary to 3-D systems we have no gravothermal catastrophe
(associated with the disappearance of any local entropy maximum, e.g. 
Binney \& Tremaine 1987; Padmanabhan 1990).

Of course, the energy $E$ is continuous at the transitions $T_{cn}$. However,
its derivative $C_v=\d E/\d T$ is discontinuous as we jump onto the
branch $\pm n$. Indeed, from eq.(\ref{E0}) the specific heat $C_{v0}$ 
associated with the uniform solution is:
\beq
C_{v0} = \frac{M}{2} ,
\label{Cv0}
\eeq
while near the transition $T_{c1}$, using the expansion 
(\ref{psik})-(\ref{a1}), we obtain:
\beq
E_1 \simeq E_{c1} + \frac{7M}{2} (T-T_{c1}) \;\;\; \mbox{hence} \;\;\;
C_{v1}(T_{c1}^-) = \frac{7M}{2} .
\label{Cv1}
\eeq
We show in fig.~\ref{figCv} the specific heat for $n=0,\pm 1$ and $\pm 2$.
We can check the discontinuity at $T_{cn}$ which corresponds to a second-order
phase transitions. Next, we display in fig.~\ref{figbetaE} the calorific curve 
$\beta(E)$. It also clearly exhibits this second-order phase transition
and we can see on the figure the various low energy bounds $E_{\rm min}(n)$.
Besides, we can see in fig.~\ref{figbetaE} that the specific heat is always 
positive and no gravothermal catastrophe can develop. Overall, the behaviour
of the OSC model is quite similar to the HMF model studied in details 
in Chavanis et al. (2005). The latter
model is defined by a cosine self-interaction $\cos[2\pi(x_i-x_j)/L]$ instead of
$|x_i-x_j|$ without the ``external'' potential $V$. It can also be seen as
a truncation of the 1-D gravitational potential $\Phi$ to its first Fourier
component. As described in Chavanis et al. (2005) the HMF model also exhibits a
second-order phase transition from an homogeneous equilibrium above $T_{c1}$
to a clustered phase with only one density peak below $T_{c1}$. However, there
are no other (unstable) equilibria $\pm n$ appearing at $T_{cn}$. Of course,
the infinite series of equilibria $\pm n$ which appear in the gravitational case
is due to the scale-free nature of the gravitational potential 
$\propto |x_i-x_j|$ (which is only broken by the finite length $L$ of the
system which leads to a discrete spectrum instead of a continuous one). We shall
check in the following that apart for the additional equilibria $\pm n$
the properties of the HMF system and of the OSC model (\ref{HN})
are similar both from thermodynamical and dynamical points of view.

\subsubsection{Canonical ensemble}
\label{Canonical-ensemble1}

\begin{figure}
\begin{center}
\epsfxsize=8 cm \epsfysize=6 cm {\epsfbox{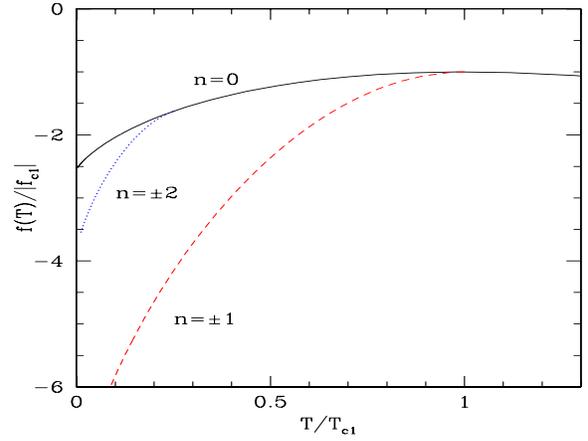}} 
\end{center}
\caption{The free energy $f(T)=F(T)+T S(T_{c1})$ as a function of temperature 
for the states $n=0,\pm 1$ and $\pm 2$.}
\label{figFT}
\end{figure}

In the canonical ensemble the thermodynamical equilibrium is obtained by
minimizing the free energy $F=E-TS$ at fixed temperature $T$ and mass $M$.
This yields again the Maxwell-Boltzmann distribution (\ref{Maxwell})
hence we recover the same critical temperatures and solutions 
(\ref{betacn})-(\ref{a1}). The free energy $F(T)$ reads:
\beqa
\lefteqn{F+T S(T_{c1}) = E-MTs = \frac{MT}{2} - \frac{gM^2L}{6}} \nonumber \\
&& - \frac{MT}{2} \ln\left(\frac{T}{T_{c1}}\right) - \frac{MT}{2} 
\int_0^{\zeta_L} \frac{\d\zeta}{\zeta_L} \; (\eta+1)\psi .
\label{F}
\eeqa
We display in fig.~\ref{figFT} the free energy $F$ as $F(T)+T S(T_{c1})$
as a function of temperature for equilibria $n=0,\pm 1$ and $\pm 2$.
We can see that at fixed $T$ the homogeneous state ($n=0$) has the highest
free energy while $F$ decreases for lower $n$ with $n\geq 1$. Therefore, as
we shall check in sect.~\ref{Canonical-ensemble2} the uniform equilibrium
is again unstable below $T_{c1}$ while the equilibrium $n=\pm 1$ is stable.

\subsubsection{Grand-canonical ensemble}
\label{Grand-canonical-ensemble1}

In the grand-canonical ensemble the thermodynamical equilibrium is obtained by
minimizing the grand potential $\Omega=E-TS-\mu M$ at fixed temperature $T$ 
and chemical potential $\mu$. Since the mass $M$ now fluctuates the mean
density $\lag\rho\rag=M/L$ within $[0,L]$ is no longer set equal to $\rhob$.
In this case, we cannot any longer extend the system to the whole real
line using a periodicity of $2L$ and symmetry with respect to $x=0$. It can
only be interpreted as a small system of actual size $L$ within a larger 
cosmological background of mean density $\rhob$, analyzed over time-scales
much smaller than the Hubble time. Thus, if $\lag\rho\rag>\rhob$ it corresponds
to a local overdensity while if $\lag\rho\rag<\rhob$ it corresponds
to a local underdensity. Then, minimizing the grand potential $\Omega$ yields
the same results (\ref{betacn})-(\ref{a1}) as for the micro-canonical and 
canonical ensembles. Note that the agreement between all three thermodynamical
ensembles does not hold for all 1-D systems. For instance a system of concentric 
spherical cells which interact by gravity can show metastable states, negative
specific heat and discrepancies between different thermodynamical
ensembles as seen in Youngkins \& Miller (2000). Thus the 1-D system studied 
here from
planar perturbations shows a rather simple behaviour while presenting a 
phase-transition and a scale-free origin.

\subsection{Thermodynamical stability}
\label{Thermodynamical-stability}

\subsubsection{Micro-canonical ensemble}
\label{Micro-canonical-ensemble2}

As recalled in sect.~\ref{Micro-canonical-ensemble1} the thermodynamical
equilibrium states within the micro-canonical ensemble are obtained by 
maximizing the entropy $S$. In sect.~\ref{Micro-canonical-ensemble1} we
investigated the first variations of the entropy $S$, which yields maxima,
minima and saddle-points. Next, the thermodynamical stability of these states
is obtained by studying the second variations of the entropy. A stable
equilibrium state corresponds to negative definite second variations.
Following Padmanabhan (1990), maximizing the entropy $S[f]$ at fixed energy $E$
and density $\rho(x)$ yields the Maxwellian distribution (\ref{Maxwell2}).
Then, we can write the entropy $S$ and the energy $E$ in terms of the density 
$\rho(x)$ as in eqs.(\ref{S2}), (\ref{E1}). Next, we can obtain from these
expressions the second variation $\delta^2S$ with respect to a density
fluctuation $\delta\rho(x)$ at fixed energy (the constraint $\delta E=0$
yields $\delta T$ in terms of $\delta\rho$, which is next substituted into 
$\delta S$). This yields for the second variation around an equilibrium 
solution:
\beqa
\lefteqn{\delta^2S = -\frac{g}{2T} \int\d x \d x' \; \delta\rho(x) 
\delta\rho(x') |x-x'| - \int \d x \; \frac{\delta\rho^2}{2\rho} } \nonumber \\
&& - \frac{1}{MT^2} \left( \int \d x \; \phi \; \delta\rho \right)^2 .
\label{d2S1}
\eeqa
As in Padmanabhan (1990) let us define the mass fluctuation $q(x)$ below $x$ 
as:
\beq
\delta\rho = \frac{\d q}{\d x} , \;\; q(x) = \int_0^x \d x' \; 
\delta\rho(x') , \;\; q(0)=q(L)=0,
\label{qrho}
\eeq
where we used the conservation of mass $M$. Then, after integrations by parts 
the second variation $\delta^2S$ can be written as:
\beq
\delta^2S = \int_0^L \d x \d x' \; q(x) K_{MC}(x,x') q(x')
\label{d2S2}
\eeq
with:
\beqa
K_{MC}(x,x') & = & \frac{g}{T} \delta_D(x-x') - \frac{1}{MT^2} 
\frac{\d\phi}{\d x}(x) \frac{\d\phi}{\d x}(x')  \nonumber \\
&& + \frac{1}{2} \delta_D(x-x') \frac{\d}{\d x'} \left( \frac{1}{\rho(x')} 
\frac{\d}{\d x'} \right) .
\label{KMC}
\eeqa
Thus, the equilibrium $\rho(x)$ is stable if all eigenvalues $\lambda$ of the
kernel $K_{MC}$ are negative (so that $S[\rho]$ is a maximum). Hence we are 
led to investigate the eigenvalue problem $K_{MC} . q = \lambda q$ which reads:
\beq
\frac{1}{2} \frac{\d}{\d x} \left[\frac{1}{\rho(x)}\frac{\d q}{\d x}\right] 
+ \frac{g}{T} q(x) = \lambda q(x) + \frac{\phi'(x)}{MT^2} \int\d x' \phi' q .
\label{qlambdaMC1}
\eeq
In terms of the dimensionless variables (\ref{psieta}) we obtain the 
eigenvalue problem:
\beq
\frac{\d}{\d\zeta} \left[\frac{1}{\eta}\frac{\d q}{\d\zeta}\right] + q =
\tilde{\lambda} q + 2 \psi' \int\frac{\d\zeta'}{\zeta_L} \psi' q ,
\;\; \mbox{with} \;\; \tilde{\lambda} = \frac{\lambda}{g\beta} ,
\label{qlambdaMC2}
\eeq
supplemented by the boundary conditions $q(0)=q(\zeta_L)=0$. We can note that
the matrix $\tilde{K}_{MC}$ defined by:
\beq
\tilde{K}_{MC} . q = - \frac{\d}{\d\zeta} \left[\frac{1}{\eta}\frac{\d q}{\d\zeta}\right] + 2 \psi' \int\frac{\d\zeta'}{\zeta_L} \psi' q 
\label{KtMC}
\eeq
is positive definite, therefore $\tilde{\lambda}\leq 1$.

\begin{figure}
\begin{center}
\epsfxsize=8 cm \epsfysize=6 cm {\epsfbox{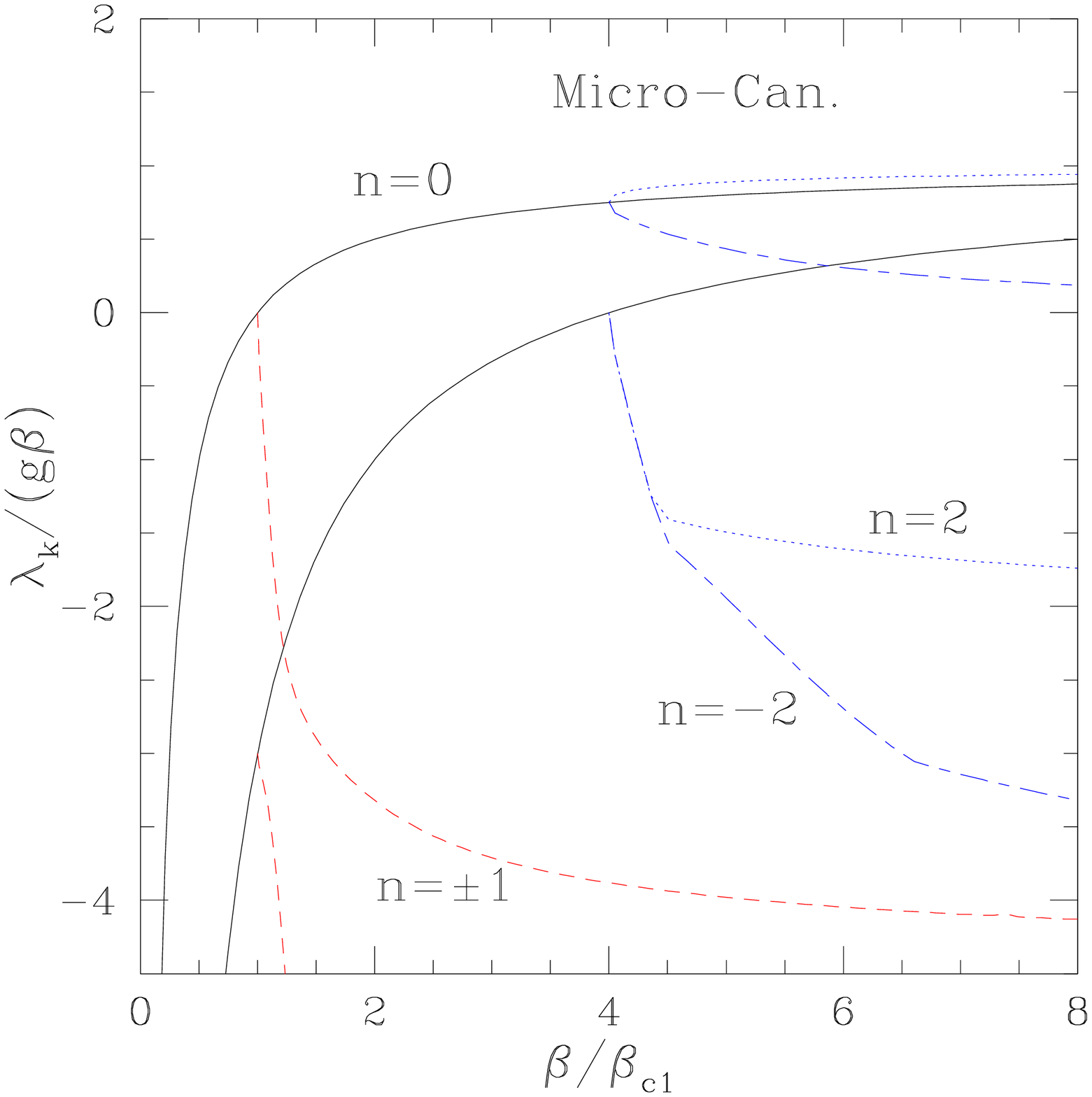}} 
\end{center}
\caption{The largest two stability eigenvalues 
$\tilde{\lambda}_k=\lambda_k/g\beta$ (i.e. $k=1,2$) as a function of inverse 
temperature $\beta$ for the equilibrium states $0$ (solid lines), $\pm 1$ 
(dashed lines), $2$ (dotted lines) and $-2$ (dot-dashed lines) for the 
micro-canonical ensemble. Stability corresponds to negative eigenvalues 
$\tilde{\lambda}_k<0$. The states $\pm 1$ are actually identical (up to a
symmetry with respect to $x=L/2$). The equilibria $\pm n$ show $n-1$ 
unstable eigenvalues while the homogeneous solution $n=0$ shows $n$ unstable 
eigenvalues below $T_{cn}$.}
\label{figlambdabetaMC}
\end{figure}

For the homogeneous equilibrium ($n=0$) where $\eta=1$ and $\psi=0$ we obtain:
\beq
q'' + (1-\tilde{\lambda}) q = 0 , \;\;\;  q(0)=q(\zeta_L)=0 ,
\label{q0MC1}
\eeq
whose solutions are:
\beq
q_k = A_k \sin\left(k\pi\frac{\zeta}{\zeta_L}\right) , \;\; 
\tilde{\lambda}_k = 1 - \left(\frac{k\pi}{\zeta_L}\right)^2 , \;\; k=1,2,..
\label{lambda0MC}
\eeq
Thus, as expected we find that the uniform equilibrium is stable at high
temperature $T>T_{c1}$ (i.e. $\zeta_L<\pi$). At each critical temperature
$T_{cn}$ defined in eq.(\ref{betacn}) a new mode of instability $\lambda_n$ 
appears while two new equilibrium states $\pm n$ appear which exhibit $n-1$ 
unstable modes. This agrees with fig.~\ref{figSE} which displays the specific 
entropy as a function of energy.
In particular, close to $T_{c1}^-$ we can investigate the
stability of equilibrium $n=\pm 1$ by using the expansion 
(\ref{psik})-(\ref{a1}) and looking for a similar expansion for $\lambda_1$. 
Thus, we write:
\beq
q_1 = q_{1,0} + a_1 q_{1,1} + a_1^2 q_{1,2} + .. , \;\;\;
\tilde{\lambda}_1 = a_1^2 \tilde{\lambda}_{1,2} 
+ .. ,
\label{lambda1MC}
\eeq
which yields (we can check that $\tilde{\lambda}_{1,1}=0$):
\beq
q_{1,0}=\sin\zeta , \; q_{1,1}= B_1 \sin\zeta - \frac{1}{3} \sin 2\zeta ,
\;\;\; \tilde{\lambda}_{1,2} = - \frac{5}{4} .
\eeq
Therefore, since $\lambda_1<0$ close to $T_{c1}^-$ we see that equilibria 
$\pm 1$ are stable close to $T_{c1}^-$ while equilibrium $0$ becomes unstable.
From fig.~\ref{figSE} we see that the states $\pm 1$ are actually stable down 
to $T=0$.
We display in fig.~\ref{figlambdabetaMC} the first two stability eigenvalues 
$\tilde{\lambda}_k$ as a function of $\beta$ for the equilibrium solutions
$n=0,\pm 1$ and $\pm 2$, obtained from a numerical study of 
eq.(\ref{qlambdaMC2}). We can check that the equilibria $\pm 1$ are stable above
$\beta_{c1}$ while the homogeneous equilibrium is unstable. Higher states
$\pm n$ exhibit $n-1$ unstable eigenvalues therefore they are only 
saddle-points of the entropy $S[f]$ and they are not thermodynamically stable.

The series of equilibrium states $\pm n$ which appear as new unstable modes
from the homogeneous solution at increasingly low $T_{cn}$ are somewhat analogous
to the high order modes of instability which appear for a 3-D finite isothermal 
sphere obtained in Semelin et al. (1999) and Chavanis (2002). They can be related 
to the scale-free nature of gravitational interactions (although the homogeneous
background yields an arithmetic hierarchy of scales while the isothermal sphere
leads to a geometric progression). However, since the equilibria $\pm n$ with
$n\geq 2$ are unstable (except $n=2$ at low $T$ for a collisionless dynamics as
seen in sect.~\ref{Vlasov} below) a discrete $N-$body system which undergoes
collisional relaxation is not expected to fragment into many clumps at low $T$
but to form only one collapsed density peak close to one boundary (though
several temporary clumps might appear in a transient regime).

\subsubsection{Canonical ensemble}
\label{Canonical-ensemble2}

\begin{figure}
\begin{center}
\epsfxsize=8 cm \epsfysize=6 cm {\epsfbox{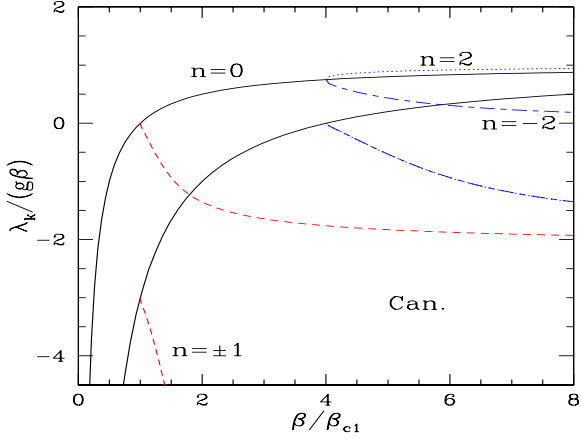}} 
\end{center}
\caption{The stability eigenvalues $\tilde{\lambda}_k=\lambda_k/g\beta$ as a 
function of inverse temperature $\beta$ for the equilibria $0,\pm 1$ and $\pm 2$ 
for the canonical ensemble. Stability corresponds to negative eigenvalues 
$\tilde{\lambda}_k<0$.}
\label{figlambdabetaC}
\end{figure}

Within the canonical ensemble we need to minimize the free energy $F$. 
Hence stable equilibria are characterized by negative definite second
variations of $-F$ at fixed temperature and mass. Following the same procedure
as in sect.~\ref{Micro-canonical-ensemble2} we obtain a similar eigenvalue
problem with a kernel $K_C(x,x')$ which is equal to $K_{MC}(x,x')$ written
in eq.(\ref{KMC}) without the term $\phi'(x)\phi'(x')/MT^2$. Therefore, the 
eigenvalue problem now reads:
\beq
\frac{\d}{\d\zeta} \left[\frac{1}{\eta}\frac{\d q}{\d\zeta}\right] + q =
\tilde{\lambda} q , \;\;\; q(0)=q(\zeta_L)=0 .
\label{qlambdaC1}
\eeq
We again obtain as in eq.(\ref{KtMC}) that $\tilde{\lambda}\leq 1$.
Then, the stability properties of the uniform solution are exactly the same
as for the micro-canonical ensemble (since $\psi=0$). On the other hand, for 
equilibrium $1$ we can again use the expansion (\ref{psik})-(\ref{a1}) close to 
$T_{c1}^-$. This now yields $\tilde{\lambda}_{1,2} = -1/4$. Therefore, 
states $\pm 1$ are again stable close to $T_{c1}^-$. 
We display in fig.~\ref{figlambdabetaC} the stability eigenvalues obtained
from a numerical study of eq.(\ref{qlambdaC1}). We obtain a behaviour
which is similar to fig.~\ref{figlambdabetaMC} and we recover the same
thermodynamical stability properties. This agrees with fig.~\ref{figFT}
for the free energy. Note that the equivalence between the micro-canonical 
and the canonical ensemble (with, however, different eigenvalues) obtained in 
this case is similar to that found for the HMF model but differs from the case 
of 3-D self-gravitating systems where the limits of stability do not coincide in
micro-canonical and canonical ensembles(Padmanabhan 1990; Chavanis 2002).
This is also related to the fact that the specific heat is always positive
(see fig.~\ref{figCv}) which is not the case for 3-D self-gravitating systems
(in the micro-canonical ensemble).

\subsubsection{Grand-canonical ensemble}
\label{Grand-canonical-ensemble2}

\begin{figure}
\begin{center}
\epsfxsize=8 cm \epsfysize=6 cm {\epsfbox{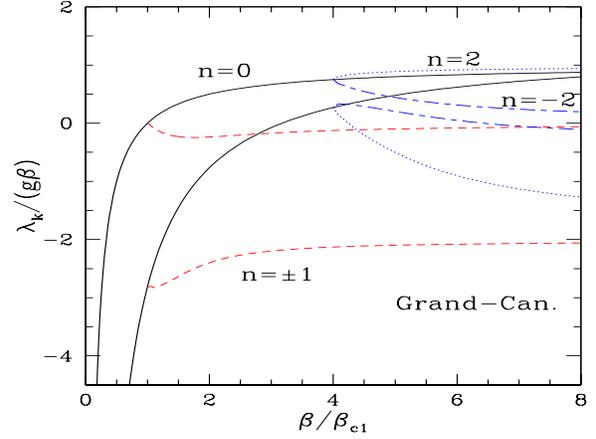}} 
\end{center}
\caption{The stability eigenvalues $\tilde{\lambda}_k=\lambda_k/g\beta$ as a 
function of inverse temperature $\beta$ for the equilibria $0,\pm 1$ and $\pm 2$ 
for the grand-canonical ensemble. Stability corresponds to negative eigenvalues 
$\tilde{\lambda}_k<0$.}
\label{figlambdabetaGC}
\end{figure}

Within the grand-canonical ensemble we now need to minimize the grand potential
$\Omega$. The second variation of $\Omega$ reads:
\beq
\delta^2\Omega = \frac{g}{2} \int\d x \d x' \; \delta\rho(x) \delta\rho(x') 
|x-x'| + T \int \d x \; \frac{\delta\rho^2}{2\rho} .
\label{d2Om1}
\eeq
Since the total mass is now allowed to fluctuate we define the perturbation
$\delta\rhob$ of the mean density $M/L$ and $q(x)$ by:
\beq
\delta\rho(x)=\delta\rhob+q'(x) , \;\;\; q(0)=q(L)=0.
\label{qrhoGC}
\eeq
Substituting into eq.(\ref{d2Om1}) and integrating by parts yields:
\beqa
\lefteqn{\delta^2\Omega = - g \int\d x \; q^2 - \frac{T}{2} \int\d x \; 
q \frac{\d}{\d x} \left(\frac{q'}{\rho}\right) + \frac{\delta\rhob^2}{2}
\frac{gL^3}{3} } \nonumber \\
&& + \frac{\delta\rhob^2}{2} T \int \frac{\d x}{\rho} 
- \delta\rhob \int\d x \; q \left[-T\frac{\rho'}{\rho^2} + g (2x-L) \right]  .
\label{d2Om2}
\eeqa
Eq.(\ref{d2Om2}) shows that $\delta^2\Omega$ has a finite minimum over 
$\delta\rhob$. Thus, we first minimize $\delta^2\Omega$ over $\delta\rhob$
and the second variation of $\Omega$ with respect to $q(x)$ alone now reads:
\beqa
\lefteqn{-\frac{\delta^2\Omega}{T}(q) = \frac{1}{2} \int\d x \; q 
\frac{\d}{\d x} \left(\frac{q'}{\rho}\right) + \frac{g}{T} \int\d x \; q^2 }
\nonumber \\
&& + \left(\frac{gL^3}{3T}+\int\frac{\d x}{\rho}\right)^{-1}
\left[ \int\d x \; q [\frac{g}{T}(x-\frac{L}{2})-\frac{\rho'}{2\rho^2} ] 
\right]^2 .
\label{d2Om3}
\eeqa
This can again be written in terms of a kernel $K_{GC}(x,x')$ which leads
to the eigenvalue problem:
\beqa
\lefteqn{\frac{\d}{\d\zeta} \left[\frac{1}{\eta}\frac{\d q}{\d\zeta}\right] 
+ q = \tilde{\lambda} q 
- \left(\frac{\zeta_L^3}{6}+\int\frac{\d\zeta}{\eta}\right)^{-1} } \nonumber \\
&& \times \left(\frac{\d}{\d\zeta}\left(\frac{1}{\eta}\right)
+(\zeta-\frac{\zeta_L}{2}) \right) \nonumber \\
&& \times \int \d\zeta' \left(\frac{\d}{\d\zeta'}\left(\frac{1}{\eta}\right)
+(\zeta'-\frac{\zeta_L}{2})\right) q(\zeta') .
\label{qGC1}
\eeqa
For the uniform solution this equation reads:
\beq
q''+ (1-\tilde{\lambda}) q = - \frac{\zeta-\frac{\zeta_L}{2}}
{\zeta_L+\frac{\zeta_L^3}{6}} \int\d\zeta' \; (\zeta'-\frac{\zeta_L}{2}) 
\; q(\zeta') .
\label{q0GC1}
\eeq
For odd $k$ we recover the modes (\ref{lambda0MC}) of the micro-canonical and
canonical ensembles (since the r.h.s. in eq.(\ref{q0GC1}) vanishes) whereas
even modes are modified. In particular, their eigenvalues are now given by:
\beq
\theta \cot\theta = 1 - \frac{\theta^2}{3} - \theta^4 \left( 
\frac{16}{\zeta_L^4} + \frac{8}{3\zeta_L^2} \right) \; \mbox{with} \;
\tilde{\lambda} = 1-\left(\frac{2\theta}{\zeta_L}\right)^2  .
\label{lambda0GC}
\eeq
For large $\theta$ this yields $\theta_{2k} \simeq k\pi$ hence we recover
the even modes $\tilde{\lambda}_{2k} \simeq 1-(2k\pi/\zeta_L)^2$ of 
(\ref{lambda0MC}). One can see from eq.(\ref{lambda0GC}) that 
$\tilde{\lambda}$ is increased at fixed $\zeta_L$ as compared with the
micro-canonical and canonical result (\ref{lambda0MC}). We can note that
for $\zeta_L > \sqrt{60+12\sqrt{30}}$ the first even mode has 
$\tilde{\lambda}_2>1$, that is $\theta_2$ is imaginary and can be written
as $\theta_2=i\hat{\theta}_2$ with $\hat{\theta}_2\coth\hat{\theta}_2 = 1 + 
\hat{\theta}_2^2/3 - \hat{\theta}_2^4 (16/\zeta_L^4 + 8/3\zeta_L^2)$.
In any case, we find that as for the micro-canonical and canonical ensembles
the homogeneous equilibrium is unstable below $T_{c1}$ while other equilibrium
solutions $\pm n$ are also unstable except for $n=1$, as can be seen from
fig.~\ref{figlambdabetaGC} where we show the results of a numerical analysis
of the eigenvalue problem (\ref{qGC1}). Again, it is possible to analyze the 
stability of state $\pm 1$ near the transition $T_{c1}$ using the expansion 
(\ref{psik})-(\ref{a1}). This yields $\tilde{\lambda}_{1,2} = -1/4<0$ which
implies that equilibria $\pm 1$ are again stable close to $T_{c1}^-$. From  
fig.~\ref{figlambdabetaGC} we see they are actually stable down to $T=0$.
Thus, all three thermodynamical ensembles lead to the same stability 
properties.

\section{Hydrodynamical model}
\label{Hydrodynamical-model}

The thermodynamical analysis of sect.~\ref{Thermodynamical-analysis}
has allowed us to obtain the equilibrium states of the OSC system (\ref{HN})
and to derive their thermodynamical stability properties. However, this
analysis does not yield the growth rates associated with such instabilities
nor the dynamical properties of the system. Within a mean field or continuum
approach we shall study in sect.~\ref{Vlasov} the dynamics as given by
the Vlasov-Poisson system. However, before we tackle the Vlasov equation
in phase-space $(x,v)$ we first investigate in this section the hydrodynamical
model associated with the hamiltonian (\ref{HN}). Indeed, the fluid and
collisionless systems often share important properties while the hydrodynamical
system is much easier to analyze (e.g. Binney \& Tremaine 1987).

\subsection{Hydrostatic equilibrium}
\label{Hydrostatic-equilibrium}

The equations of motion of the hydrodynamical system associated with the
OSC model (\ref{HN}) are simply:
\beq
\frac{\pl\rho}{\pl t} + \frac{\pl}{\pl x} (\rho v) = 0 ,
\label{Hydromass}
\eeq
\beq
\frac{\pl v}{\pl t} + v \frac{\pl v}{\pl x} = - \frac{1}{\rho} 
\frac{\pl P}{\pl x} - \frac{\pl \phi}{\pl x} ,
\label{Euler}
\eeq
where the potential $\phi$ is given by eq.(\ref{phi2}) and $P(\rho)$ is the 
barotropic pressure of the gas. As usual, one can check that the energy 
functional $W$ is conserved by the equations of motion (Binney \& Tremaine 1987),
with:
\beq
W= \int\d x \; \rho(x) \left[ \int_0^{\rho} \d\rho' \frac{P(\rho')}{\rho'^2}
\right] + \int\d x \; \rho \frac{v^2+\phi+V}{2} .
\label{Whydro}
\eeq
Therefore, minima of $W$ are stationary solutions of the Euler equations
which are non-linearly stable. As in the statistical analysis of 
sect.~\ref{Statistical-equilibrium} the first variation of $W$ yields the
properties of such stationary states (in addition to saddle-points of $W$).
Then, the first variation of $W$ with respect to $v$ leads to $v=0$. 
Substituting into the Euler eq.(\ref{Euler}) yields the hydrostatic equilibrium
condition:
\beq
\frac{\d P}{\d x} = - \rho \frac{\d\phi}{\d x} .
\label{hydrostatic}
\eeq
In the following we shall restrict ourselves to the case of the isothermal gas 
defined by the equation of state $P=\rho T$. Then eq.(\ref{hydrostatic})
reads $\d\rho/\d x= - \beta \rho \d\phi/\d x$ which yields again 
$\rho\propto e^{-\beta \phi}$ as in the thermodynamical approach 
(\ref{Maxwell}). Therefore, we recover the same stationary states as in
sect.~\ref{Statistical-equilibrium}.

\subsection{Non-linear stability}
\label{Non-linear-stability}

The equilibrium states obtained in sect.~\ref{Hydrostatic-equilibrium} are
non-linearly stable if they are minima of the energy functional $W$, that is
if the second variations of $W$ are definite positive. The second variation 
with respect to $v$ is clearly positive from the expression (\ref{Whydro})
while $\delta^2W/\delta\rho\delta v=0$ since $v=0$ for equilibrium states.
Therefore, we only need to study the second variation with respect to the 
density. For the isothermal gas this yields:
\beq
\delta^2 W= \frac{1}{2}\int\d x \; \delta\rho\delta\Phi + \frac{T}{2}
\int\d x \; \frac{\delta\rho^2}{\rho} .
\label{d2W}
\eeq
This is the same equation as the second variation of the free energy $F$
obtained for the canonical ensemble in sect.~\ref{Canonical-ensemble1}. Hence
the non-linear stability properties of various equilibrium states are the 
same as those derived for the canonical ensemble. Therefore, the non-linear
hydrodynamical stability properties are identical to the thermodynamical
ones. This implies in particular from eq.(\ref{lambda0MC}) that the homogeneous 
equilibrium is unstable below $T_{c1}$ while equilibria $\pm n$ are unstable 
close to $T_{cn}$ for $n\geq 2$. From fig.~\ref{figlambdabetaC} they remain
unstable down to $T\rightarrow 0$ while equilibrium $\pm 1$ is stable below
$T_{c1}$.

\subsection{Dynamical linear stability}
\label{Dynamical-linear-stability}

\begin{figure}
\begin{center}
\epsfxsize=8 cm \epsfysize=6 cm {\epsfbox{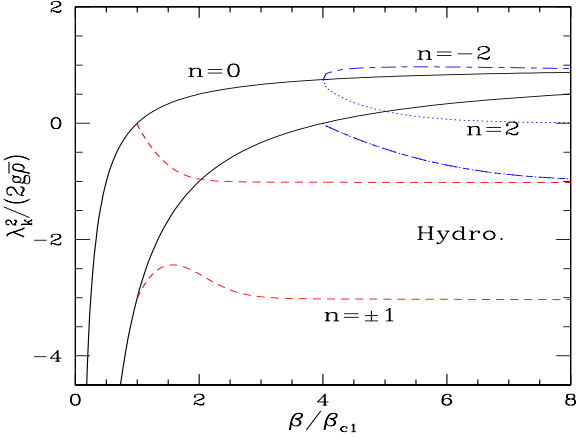}} 
\end{center}
\caption{The stability eigenvalues $\nu_k=\lambda_k^2/(2g\rhob)$ as a 
function of inverse temperature $\beta$ for the equilibria $0,\pm 1$ and 
$\pm 2$ for the hydrodynamical system. Stability corresponds to imaginary 
eigenvalues $\lambda_k^2<0$. The second eigenvalues $\lambda_2$ obtained for
equilibria $\pm 2$ are equal.}
\label{figlambdabetaHydro}
\end{figure}

The analysis of sect.~\ref{Non-linear-stability} allowed us to discriminate 
between stable and unstable states but it does not provide the growth rates
of unstable perturbations. To this order we need to investigate the dynamical
linear stability of such equilibrium states. Thus, we linearize the equations
of motion (\ref{Hydromass})-(\ref{Euler}) and we study the normal modes of
the form $\delta\rho \sim e^{\lambda t}$. Introducing again the mass 
fluctuation $q(x)$ as in eq.(\ref{qrho}) and using the dimensionless
variables (\ref{psieta}) we obtain the eigenvalue problem:
\beq
\frac{\d}{\d\zeta} \left[\frac{1}{\eta}\frac{\d q}{\d\zeta}\right] + q =
\nu \frac{q}{\eta} \;\; \mbox{with} \;\; \nu= \frac{\lambda^2}{2g\rhob}
\;\; \mbox{and} \;\; \delta v=-\lambda \frac{q}{\rho} .
\label{qlambdahydro}
\eeq
Thus, eigenmodes with $\nu<0$ (i.e. $\lambda$ is imaginary) are stable
while eigenmodes with $\nu>0$ (i.e. $\lambda$ is real) are unstable.
More precisely, for $\nu>0$ we have both a decaying and a growing mode
($\lambda=\pm \sqrt{\nu 2g\rhob}$) distinguished by opposite velocity fields,
as seen from the last equation in (\ref{qlambdahydro}). 
At the point of marginal stability $\nu=0$ we 
recover the eigenvalue problem (\ref{qlambdaC1}) of the canonical ensemble. 
Therefore, from sect.~\ref{Non-linear-stability} we see that the conditions 
of non-linear and linear stability are identical and also coincide with the 
canonical ensemble. Note that this equivalence also holds for the HMF model
(Chavanis et al. 2005) as well as for some more general models 
(Chavanis 2006).

For the uniform equilibrium ($n=0$) we have $\eta=1$ so that we recover the
eigenvalue problem (\ref{q0MC1}) of both the micro-canonical and canonical 
ensembles with the same solutions (\ref{lambda0MC}). Thus we find again that
the homogeneous solution is unstable below $T_{c1}$. For the equilibrium state
$\pm 1$ we can again use the expansion (\ref{psik})-(\ref{a1}) close to 
$T_{c1}$ which yields $\nu_1 \simeq -a_1^2/4<0$. Therefore the equilibrium
$\pm 1$ is again stable close to $T_{c1}$. From fig.~\ref{figlambdabetaHydro}
which shows the eigenvalues $\nu_k$ as a function of inverse temperature
$\beta$ we can check that the equilibrium $\pm 1$ is stable below $T_{c1}$ 
down to $T=0$ whereas equilibria $\pm n$ with $n\geq 2$ are always unstable.
This is identical to the non-linear stability properties obtained in 
sect.~\ref{Non-linear-stability} and to the thermodynamical results of
sect.~\ref{Thermodynamical-stability}. We can note that the equilibria 
$\pm 2$ have different first eigenvalues $\nu_1>0$ but identical second
eigenvalues $\nu_2<0$. Indeed, we can check numerically that this eigenmode
vanishes both at the boundaries $x=0,L$, and at the midpoint $L/2$
($q$ is antisymmetric with respect to $L/2$) so that both $\eta$ and 
$\delta\eta$ are identical for $n=2$ and $n=-2$ if we extend them to the whole 
real axis up to a mere translation of $L/2$. This is due to the symmetry
properties of equilibrium profiles $\pm 2$.

\begin{figure}
\begin{center}
\epsfxsize=8 cm \epsfysize=6 cm {\epsfbox{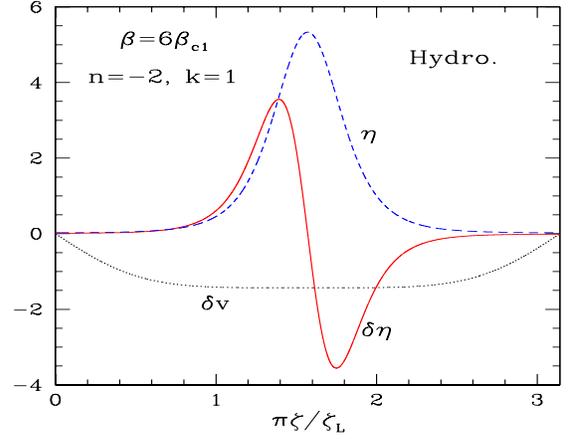}} 
\end{center}
\caption{The unstable eigenmode $k=1$ with $\nu_1 \simeq 1$ for the 
equilibrium state $n=-2$ at temperature $\beta=6\beta_{c1}$. We plot the 
equilibrium density profile $\eta=\rho/\rhob$ (dashed line), the unstable 
eigenmode $\delta\eta=\delta\rho/\rhob$ (solid line) and the velocity 
$\delta v \propto -q/\eta$ (dotted line). This eigenmode corresponds to a mere 
translation of the density peak. The normalizations of $\delta\eta$ and 
$\delta v$ are arbitrary.}
\label{figqHydrom2}
\end{figure}

Next, we can note that the eigenvalue $\nu_1$ obtained for the state
$n=-2$ quickly becomes very close to unity above $T_{c2}$. In fact, we can
check that $q=\eta$ is a solution of the differential eq.(\ref{qlambdahydro})
with $\nu=1$, using $\eta=e^{-\psi}$ and $\psi''=\eta-1$ which characterize
the equilibrium profiles, see eqs.(\ref{psieta}) and (\ref{LE2}).
Because of the boundary conditions $q(0)=q(\zeta_L)=0$ this is not an
allowed solution to the eigenmode problem (\ref{qlambdahydro}). However, for
the equilibrium $n=-2$ where the overdensity goes to zero at the boundaries
as $\eta\sim e^{-\pi^2 T_{c2}/2T}$ from eq.(\ref{etavoid}) the unstable
eigenmode tends to $q\propto \eta$ at low $T$. From the definition (\ref{qrho})
this implies for the density perturbation $\delta\eta \propto \eta'(\zeta)$
which corresponds to a mere translation of the equilibrium density profile.
We display in fig.~\ref{figqHydrom2} the first eigenmode $\nu_1\simeq 1$ 
for the equilibrium state $n=-2$ at temperature $T_{c2}/6$. We plot the
equilibrium profile $\eta$ (dashed line) and the growing mode $\delta\eta$
(solid line) with the associated velocity $\delta v$ (dotted line). 
The derivative $\eta'(\zeta)$ normalized to the $\delta\eta$
at its peak cannot be distinguished from the perturbation $\delta\eta$ on the
figure and we can see that the velocity is almost constant over the extent of
the density peak. Thus the unstable mode obtained for the equilibrium $n=-2$ 
only corresponds to a translation of the density peak (which behaves as an 
isolated system as discussed in sect.~\ref{Micro-canonical-ensemble1}) while 
other eigenmodes $k\geq 2$ are decaying modes, as seen from 
fig.~\ref{figlambdabetaHydro}. The associated eigenvalue $\nu_1=1$ merely 
corresponds to a point-particle which rolls down the ``external'' potential
$V(x)$ of eq.(\ref{phi2}). Indeed, the solution of $\ddot x= - \d V/\d x$
reads $(x-L/2) \propto e^{\pm \sqrt{2g\rhob} t}$, which describes the 
motion in the low-temperature limit $T\rightarrow 0$ as the density peak 
(\ref{etacosh}) tends to a Dirac. It would be interesting to follow the 
dynamics further from $x=L/2$ to see whether the reflection at the boundaries
$x=0,L$, quickly destroys the equilibrium profiles or whether some stationary 
cycle can be reached (if there were true solitonic solutions). We shall leave
this analysis for future work as it goes beyond the scope of the present paper.

\begin{figure}
\begin{center}
\epsfxsize=8 cm \epsfysize=6 cm {\epsfbox{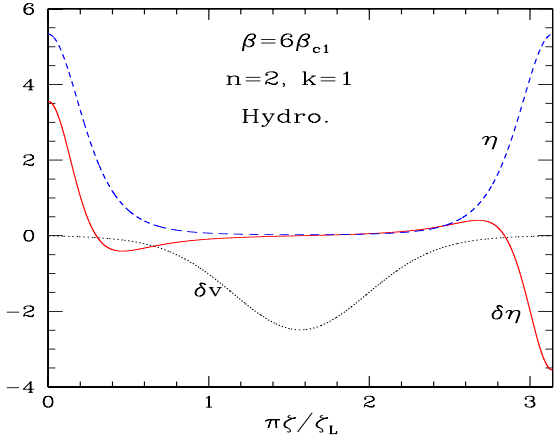}} 
\end{center}
\caption{The unstable eigenmode $k=1$ with $\nu_1>0$ for the equilibrium 
state $n=2$ at temperature $\beta=6\beta_{c1}$. We plot the equilibrium 
profile $\eta$ (dashed line), the unstable eigenmode $\delta\eta$ (solid line) 
and the velocity $\delta v$ (dotted line).}
\label{figqHydro2}
\end{figure}

Finally, we can see in fig.~\ref{figlambdabetaHydro} that for the equilibrium 
$n=2$ the eigenvalue $\nu_1$ quickly goes to zero. Of course, we cannot 
recover the solution $q\propto\eta$ associated with a mere translation since
the state $n=2$ is made of two density peaks at $x=0,L$, which grow at low $T$
so that the boundary conditions $q(0)=q(L)=0$ forbid the solution 
$q\propto\eta$. We show in fig.~\ref{figqHydro2} the first eigenmode 
$\nu_1\ga 0$ for the equilibrium state $n=2$. It corresponds to a transfer of 
matter from one density peak towards the other one. The end-product of this
process will be the equilibrium state $\pm 1$ which is the unique non-linearly
stable equilibrium, as seen in sect.~\ref{Non-linear-stability}. The growth
rate of this instability quickly goes to zero at low $T$ as the underdense 
region which connects both density peaks reaches very low densities from
eq.(\ref{etavoid}) and the density peaks behave as almost isolated subsystems.
Besides, it can be seen from the shape of $\delta\eta$ in 
fig.~\ref{figlambdabetaHydro} that as expected $\eta+\delta\eta$ corresponds
to two unequal density peaks at $x=0,L$, which follow the profiles associated
with isolated relaxed density peaks. Thus, in agreement with (\ref{etacosh})
we find that in the case displayed in fig.~\ref{figlambdabetaHydro} the peak
at $x=0$ which gains some mass gets a higher density maximum and a smaller
extension while the opposite holds for the peak at $x=L$ which looses mass.
This can be checked more precisely from eq.(\ref{qlambdahydro}) as follows.
The modified Lane-Emden eq.(\ref{LE1}) can also be written as:
\beq
\frac{\d^2}{\d\zeta^2}(-\ln\eta)=\eta-1 , \;\; \mbox{whence} \;\; 
\frac{\eta''}{\eta} - \frac{\eta'^2}{\eta^2} + \eta =1 ,
\label{LE3}
\eeq
which implies that if both $\eta$ and $\eta+\delta\eta$ are isothermal 
equilibrium solutions the difference $\delta\eta$ satisfies at first order:
\beq
\frac{\delta\eta''}{\eta} - 2 \frac{\eta'}{\eta^2} \delta\eta' 
+ \left(1-\frac{\eta''}{\eta^2}+2\frac{\eta'^2}{\eta^3}\right) \delta\eta = 0 .
\label{LEdeltaeta}
\eeq
On the other hand, taking the derivative of the eigenmode equation 
(\ref{qlambdahydro}) yields (with $q'=\delta\eta$):
\beq
\frac{\delta\eta''}{\eta} - 2 \frac{\eta'}{\eta^2} \delta\eta' 
+ \left(1-\frac{\eta''}{\eta^2}+2\frac{\eta'^2}{\eta^3}\right) \delta\eta = 
\nu \left( \frac{\delta\eta}{\eta} - \frac{\eta'}{\eta^2} q \right) .
\label{etalambdahydro}
\eeq
Therefore, for $\nu=0$ we recover eq.(\ref{LEdeltaeta}). In fact, it is clear
that if $(\rho_1,v_1)$ and $(\rho_2,v_2)$ are two equilibrium solutions of the
hydrodynamic eqs.(\ref{Hydromass})-(\ref{Euler}) the difference 
$(\rho_2-\rho_1,v_2-v_1)$ is a static solution at linear order of the linearized
equations of motion about $(\rho_1,v_1)$. Thus, in agreement with
fig.~\ref{figlambdabetaHydro} the unstable eigenmode $\nu_1\rightarrow 0^+$ 
obtained at low temperature for the equilibrium $n=2$ actually yields a 
quasi-static transfer of matter between the two density peaks in the sense
that they follow equilibrium profiles (parameterized by their varying mass) at 
the fixed temperature $T$.

It can be checked by a numerical computation of the eigenmode problem 
(\ref{qlambdahydro}) that the behaviours obtained for equilibria $\pm 2$ extend 
in a natural manner to higher-order equilibrium states $\pm n$.
Thus, equilibria $\pm n$ show $n-1$ unstable eigenmodes (as for the non-linear
stability and thermodynamical stability analysis) which can be understood as
translations of the density peaks
or as slow transfers of matter between neighbouring density peaks.
For instance, the equilibrium $n=4$ which corresponds to three density peaks
at $x=0,L/2$ and $L$ exhibits one eigenvalue $\nu_1 \simeq 1$ and two positive 
eigenvalues $\nu_2>\nu_3$ close to zero. For the mode $\nu_1$ the central
peak rolls to the left or to the right. For the mode $\nu_2$ which is symmetric
with respect to $x=L/2$ there is a transfer of matter from the
central peak towards both left and right peaks (or the opposite since $-q$ is
also a growing eigenmode). For the mode $\nu_3$ which is antisymmetric
with respect to $x=L/2$ (for $\delta\eta$) there is a transfer of matter from
the right peak to the central one and a second transfer with the same amplitude
from the central peak to the left one (i.e. a flow from the right to the left
or the opposite for $-q$). On the other hand, the equilibrium $n=-4$ which 
corresponds to two density peaks at $x=L/4$ and $3L/4$ exhibits two eigenvalues 
$\nu_1>\nu_2$ close to unity and one positive eigenvalue $\nu_3$ close to zero.
For the mode $\nu_1$ both peaks move with the same velocity towards the
same direction while for mode $\nu_2$ they move in opposite directions.
The mode $\nu_3$ involves a transfer of matter between both peaks.
Note that the largest growing rate corresponds to both peaks moving together
in the same direction: the system is not analogous to two particles in the
external potential $V$ alone as one needs to take into account the gravitational
interaction $\Phi$ which balances $V$ at equilibrium. We would obtain similar
results for higher $n$. As noticed above, it would be interesting to follow
the evolution of the system by numerical simulations to study the possible
merging of these density peaks.

\section{Vlasov equation}
\label{Vlasov}

As discussed in sect.~\ref{cosmological-gravitational-system} in the mean field 
limit where the mass of particles goes to zero the $N-$body system (\ref{HN}) 
is actually described by the Vlasov-Poisson system similar to (\ref{Vlasov2}). 
As is well-known, the equilibrium states obtained in the hydrodynamical 
approach of sect.~\ref{Hydrodynamical-model} are also stationary solutions of the
Vlasov-Poisson system with the Maxwell-Boltzmann distribution $f(x,v)$
of eq.(\ref{Maxwell}). Therefore, we recover the equilibrium profiles
described in sect.~\ref{Equilibrium-profiles:}. 

On the other hand, as described in Chavanis (2003) and Tremaine et al. (1986),
the Vlasov dynamics conserves any quantity of the form 
$H[f]=-\int C(f) \d x \d v$ so that maxima of any functional $H$ where $C(f)$
is convex are non-linearly stable. A particular case of $H[f]$ is the entropy
$S[f]$ as given in eq.(\ref{S1}) hence we can already infer that the equilibrium
$\pm 1$ which is a maximum of $S$ is stable below $T_{c1}$. However, this 
criterion cannot decide whether other equilibria $\pm n$ with $n\geq 2$ are 
stable or not.

Next, in order to study the
dynamical linear stability of these equilibrium solutions we linearize the
equations of motion. We first introduce the action-angle variables $(J,w)$
which describe the motion of a particle of energy $E$ along its orbit in
the equilibrium potential $\phi_0(x)$ from position $x_-$ to $x_+$
(see Fridman \& Polyachenko 1984; Polyachenko \& Shukhman 1981):
\beq
J= \frac{2}{2\pi} \int_{x_-}^{x_+}\d x \sqrt{2(E-\phi_0(x))} \; , \;\; 
w= \frac{\pl S}{\pl J} ,
\label{Jw}
\eeq
with:
\beq
S(J,x) = \int_{x_-}^x \d x' \sqrt{2(E-\phi_0(x'))} .
\label{S}
\eeq
The action-angle variables $(J,w)$ obey Hamilton's equations:
\beq
\dot{w} = \frac{\pl H}{\pl J} = \Omega , \;\;\; 
\dot{J} = - \frac{\pl H}{\pl w} = 0 ,
\label{Hamilton}
\eeq
while the Vlasov equation reads:
\beq
\frac{\d f}{\d t} = \frac{\pl f}{\pl t} + \dot{w} \frac{\pl f}{\pl w} 
+ \dot{J} \frac{\pl f}{\pl J} = 0 .
\label{VlasovJw0}
\eeq
We write the distribution function as $f=f_0(J)+f_1(J,w)$ where 
$f_0(J)=f_0(E)$ is the equilibrium solution and $f_1$ the perturbation. Then,
the linearized Vlasov equation for $f_1$ writes:
\beq
\frac{\pl f_1}{\pl t} + \Omega_0 \frac{\pl f_1}{\pl w} = 
\frac{\pl \phi_1}{\pl w} \frac{\d f_0}{\d J} ,
\label{VlasovJw1}
\eeq
where $\phi_1$ is the perturbed potential. Because of the periodicity of $2\pi$
with respect to $w$ of the motion of the particles we can write $\phi_1$ as:
\beq
\phi_1(J,w,t) = \frac{1}{2\pi} \sum_{l=-\infty}^{\infty} \tilde{\phi}_{1l}(J) 
e^{-i(\omega t - l w)} ,
\label{phi1l}
\eeq
where we look for eigenfrequencies $\omega$. Substituting into 
eq.(\ref{VlasovJw1}) yields:
\beq
\tilde{f}_1 = - \frac{1}{2\pi} f_0'(J) \sum_l \tilde{\phi}_{1l} 
\frac{l}{\omega-l\Omega_0} e^{i l w} ,
\label{f1phi1}
\eeq
with $f_1=\tilde{f}_1 e^{-i\omega t}$. Therefore, Poisson's equation reads:
\beq
\frac{1}{2g} \frac{\d^2\tilde{\phi}_1}{\d x^2} = \tilde{\rho}_1 = 
- \frac{1}{2\pi} \int_{-\infty}^{\infty} \d v \; f_0'(J) \sum_l 
\tilde{\phi}_{1l} \frac{l e^{i l w}}{\omega-l\Omega_0} .
\label{Poissonphi1}
\eeq

\begin{figure}
\begin{center}
\epsfxsize=8 cm \epsfysize=6 cm {\epsfbox{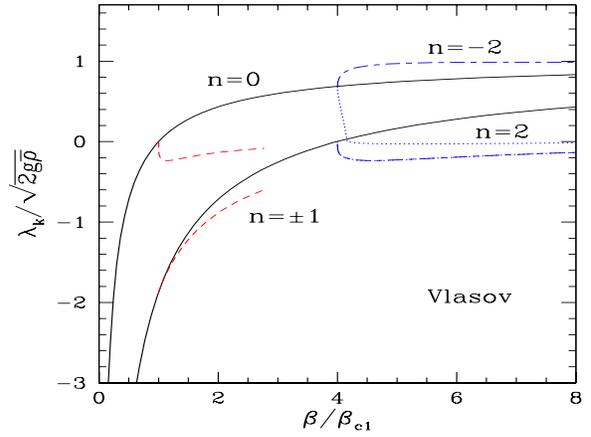}} 
\end{center}
\caption{The stability eigenvalues $\omega_k/\sqrt{2g\rhob}=
i\lambda_k/\sqrt{2g\rhob}$ as a function of inverse temperature $\beta$ for the 
equilibria $0,\pm 1$ and $\pm 2$ for the Vlasov equation of motion. Stability 
corresponds to $\lambda_k<0$. The second eigenvalues $\lambda_2$ obtained for
equilibria $\pm 2$ are equal.}
\label{figlambdabetaVlasov}
\end{figure}

In order to solve the eigenvalue problem (\ref{Poissonphi1}) it is convenient
to project this equation onto a biorthonormalized basis 
$\{\rho_{\alpha},\phi_{\alpha}\}$ such that:
\beq
\frac{\d^2\phi_{\alpha}}{\d x^2} = 2 g \rho_{\alpha} \;\; \mbox{and} \;\;
\int_0^L\d x \; \rho_{\alpha} \phi_{\alpha'} = - \delta_{\alpha,\alpha'} ,
\label{basis0}
\eeq
where $\delta_{\alpha,\alpha'}$ is the usual Kronecker symbol (Fridman \& 
Polyachenko 1984; Kalnajs 1977). For instance, we can choose with $\alpha\geq 1$:
\beq
\rho_{\alpha} = \frac{\alpha\pi}{\sqrt{gL^3}} 
\cos\left(\alpha\pi\frac{x}{L}\right) , \;\;\;
\phi_{\alpha} = - 2 \frac{\sqrt{gL}}{\alpha\pi}  
\cos\left(\alpha\pi\frac{x}{L}\right) ,
\label{basis1}
\eeq
which also satisfy the boundary conditions $\phi'(0)=\phi'(L)=0$.
Thus, if we decompose $\tilde{\phi}_1$ and $\tilde{\rho}_1$ as 
$\tilde{\phi}_1=\sum_{\alpha} a_{\alpha}\phi_{\alpha}$ and 
$\tilde{\rho}_1=\sum_{\alpha} a_{\alpha}\rho_{\alpha}$, substituting
into eq.(\ref{Poissonphi1}) and integrating over $x$ with the weight 
$\phi_{\alpha}$ yields:
\beq
a_{\alpha} = \int \frac{\d x \d v}{2\pi} \phi_{\alpha}(x) f_0'(J)
\sum_{l=-\infty}^{\infty} \sum_{\alpha'=1}^{\infty} a_{\alpha'} 
(\phi_{\alpha'})_l \frac{l e^{i l w}}{\omega-l\Omega_0}  .
\label{a-alpha}
\eeq
Using the canonical change of variable $\d x\d v= \d J\d w$ and next changing 
from action $J$ to energy $E$ we obtain for an isothermal equilibrium 
distribution $f_0 \propto e^{-\beta E}$ the eigenvalue problem:
\beq
\det ( M_{\alpha\alpha'}(\omega) - \delta_{\alpha,\alpha'} ) = 0 ,
\label{det1}
\eeq
with:
\beq
M_{\alpha\alpha'}(\omega) = - \frac{\beta}{2\pi} \int \d E f_0(E) 
\sum_l \frac{l}{\omega-l\Omega_0} (\phi_{\alpha})_l (\phi_{\alpha'})_l .
\label{Maa1}
\eeq
In terms of the dimensionless variables (\ref{psieta}) we obtain for the
basis (\ref{basis1}):
\beq
M_{\alpha\alpha'} = \frac{4\beta}{\sqrt{\pi}\beta_{c1}} 
\int_{\psi_-}^{\infty} \d\epsilon \; e^{-\epsilon} \tilde{\omega}_0 
\sum_{l=1}^{\infty} 
\frac{l^2\tilde{\omega}_0^2}{l^2\tilde{\omega}_0^2-\tilde{\omega}^2} 
(\hat{\phi}_{\alpha})_l (\hat{\phi}_{\alpha'})_l ,
\label{Maa2}
\eeq
where $\psi_-$ is the minimum of $\psi(\zeta)$ and with:
\beq
(\hat{\phi}_{\alpha})_l = \int_{\zeta_-}^{\zeta_+} \frac{\d\zeta}{\zeta_L} \;
\frac{\cos(\alpha\pi \zeta/\zeta_L) \cos(l w_+)}{\alpha\sqrt{\epsilon-\psi}} .
\label{phialpha}
\eeq
Here we introduced the reduced energy $\epsilon=\beta v^2/2+\psi$ of the 
particle, the frequencies $\tilde{\omega}_0$ and $\tilde{\omega}$ defined
by:
\beq
\tilde{\omega} =  \sqrt{\frac{\beta}{2}} \frac{L}{\pi} \omega = 
\sqrt{\frac{\beta}{\beta_{c1}}} \frac{\omega}{\sqrt{4g\rhob}}
, \;\;\; \tilde{\omega}_0 = \sqrt{\frac{\beta}{\beta_{c1}}} 
\frac{\Omega_0}{\sqrt{4g\rhob}} ,
\label{omegatilde}
\eeq
and the angular variable $w_+$ which describes the orbit in the direction
of increasing $\zeta$ from $\zeta_-$ up to $\zeta_+$:
\beq
\frac{1}{\tilde{\omega}_0}= \int_{\zeta_-}^{\zeta_+} \frac{\d\zeta}{\zeta_L} \;
\frac{1}{\sqrt{\epsilon-\psi}} , \;\;\; w_+= \pi \tilde{\omega}_0 
\int_{\zeta_-}^{\zeta} \frac{\d\zeta'}{\zeta_L} \; 
\frac{1}{\sqrt{\epsilon-\psi}} .
\label{w+}
\eeq
For the homogeneous solution we simply have $\psi=0$, 
$\tilde{\omega}_0=\sqrt{\epsilon}$, $w_+=\pi\zeta/\zeta_L$ and 
$(\hat{\phi}_{\alpha})_l = \delta_{\alpha,l}/(2\alpha\sqrt{\epsilon})$.
Making the change of variable $x=\sqrt{2\epsilon}$ we obtain:
\beq
M_{\alpha\alpha'}^0 = \delta_{\alpha,\alpha'} \frac{\beta}{\beta_{c1}} 
\frac{1}{\alpha^2\sqrt{2\pi}} \int_{-\infty}^{\infty} \d x \; e^{-x^2/2}
\; \frac{x}{x-\tilde{\omega}\sqrt{2}/\alpha} .
\label{Maa0}
\eeq
Thus, we can see from eqs.(\ref{det1}), (\ref{Maa0}), that the condition of 
marginal stability $\omega=0$ yields again $\beta=\alpha^2 \beta_{c1}$ with
$\alpha=1,2,..,$ which agrees with the critical temperatures (\ref{betacn})
obtained in the thermodynamical and hydrodynamical approaches. The matrix
$M_{\alpha\alpha'}^0$ as written above is actually defined for 
${\rm Im}(\omega)>0$, see eq.(\ref{phi1l}), which corresponds to growing modes 
(instabilities) that vanish for $t\rightarrow -\infty$. For ${\rm Im}(\omega)<0$ 
we need to deform the integration path in eq.(\ref{Maa0})
following the usual Landau analysis (e.g. Binney \& Tremaine 1987; 
Ichimaru 1973). This leads to an additional contribution
due to the residue at $x=\tilde{\omega}\sqrt{2}/\alpha$. Thus, for 
${\rm Im}(\omega)<0$ the matrix $M_{\alpha\alpha'}^0$ is given by the sum of the
expression (\ref{Maa0}) and of the contribution of the residue which reads:
\beq
{\rm Im}(\omega)<0 : \;\; {\rm Res}(M_{\alpha\alpha'}^0) = \delta_{\alpha,\alpha'}
\frac{\beta}{\beta_{c1}} \frac{2\sqrt{\pi}}{\alpha^3} i \tilde{\omega}
e^{-\tilde{\omega}^2/\alpha^2} .
\label{Res0}
\eeq
For the inhomogeneous equilibrium solutions we also need to modify the 
eq.(\ref{Maa2}) in a similar fashion for ${\rm Im}(\omega)<0$. However,
since we have no analytical expression for the integrand in eq.(\ref{Maa2})
this is more difficult. Thus, for numerical purposes we simply add the
contribution (\ref{Res0}) which is correct close to the critical points.
Therefore, the curves in fig.~\ref{figlambdabetaVlasov} are not correct for
$\lambda_k<0$ far from the transition $T_{cn}$ but the sign of $\lambda_k$
remains exact.
Following Kalnajs (1977), in order to improve the convergence with $l$ of the 
series in eq.(\ref{Maa2}) we can write 
$l^2\tilde{\omega}_0^2/(l^2\tilde{\omega}_0^2-\tilde{\omega}^2)=
1+\tilde{\omega}^2/(l^2\tilde{\omega}_0^2-\tilde{\omega}^2)$ which leads to
\beq
M_{\alpha\alpha'}(\tilde{\omega})= A_{\alpha\alpha'}(\tilde{\lambda})+
B_{\alpha\alpha'} , \;\;\; \mbox{where} \;\;\;
\tilde{\omega} = i \tilde{\lambda} ,
\label{MAB}
\eeq
with:
\beq
A_{\alpha\alpha'} = \frac{-4\beta}{\sqrt{\pi}\beta_{c1}} 
\int_{\psi_-}^{\infty} \d\epsilon \; e^{-\epsilon} \tilde{\omega}_0 
\sum_{l=0}^{\infty} 
\frac{\tilde{\lambda}^2}{\tilde{\lambda}^2+l^2\tilde{\omega}_0^2}
\frac{(\hat{\phi}_{\alpha})_l (\hat{\phi}_{\alpha'})_l}{1+\delta_{l,0}} ,
\label{A1}
\eeq
and:
\beq
B_{\alpha\alpha'} = \frac{2\beta}{\alpha\alpha'\beta_{c1}}
\int \frac{\d\zeta}{\zeta_L} \eta(\zeta)
\cos\left(\alpha\pi\frac{\zeta}{\zeta_L}\right)
\cos\left(\alpha'\pi\frac{\zeta}{\zeta_L}\right) .
\label{B1}
\eeq
In eq.(\ref{A1}) the Kronecker factor $\delta_{l,0}$ reduces the term
$l=0$ by a factor 2 as compared with $l\neq 0$. The faster convergence as 
$1/l^2$ of the expression (\ref{A1}) is convenient for numerical purposes
as it reduces the number of terms to be computed. On the other hand, the 
constant part $B_{\alpha\alpha'}$, which does not depend on $\tilde{\lambda}$,
is computed separately. Note that $A_{\alpha\alpha'}$ does not vanish for 
$\tilde{\lambda}=0$ because of the term at $l=0$. In practice, in order to
solve eq.(\ref{det1}) we do not need to compute the determinant of the matrix
$M_{\alpha\alpha'}$. Starting from the transition temperature $T_{cn}$,
where the equilibrium solution $\pm n$ which we are interested in separates
from the homogeneous equilibrium state, we follow the eigenvalue $\nu_k$
of order $k$ of $M_{\alpha\alpha'}$ and the growth rate $\tilde{\lambda}_k$
of the associated eigenmode is set by the condition 
$\nu_k(\tilde{\lambda}_k)=1$ (e.g. Kalnajs 1977). Moreover, contrary to the usual
case in plasma physics the eigenfrequencies $\omega$ are imaginary so that
$\tilde{\lambda}$ is real. This corresponds to purely growing or decaying modes
without oscillatory behaviours. Thus the matrix 
$M_{\alpha\alpha'}(\tilde{\lambda})$ is real and symmetric so that all its 
eigenvalues $\nu_k$ are real, which simplifies the computation of the roots
of $\nu_k(\tilde{\lambda}_k)=1$.

\begin{figure}
\begin{center}
\epsfxsize=8 cm \epsfysize=6 cm {\epsfbox{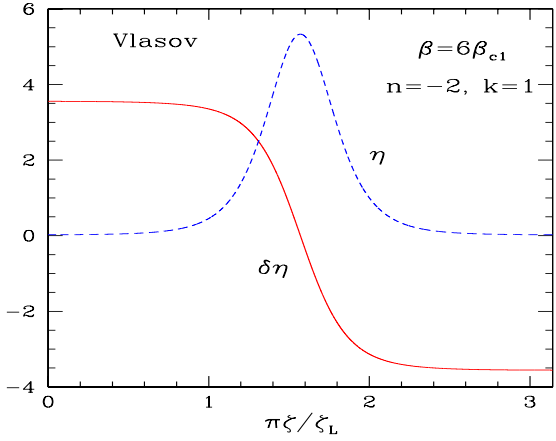}} 
\end{center}
\caption{The eigenmode $k=1$ for the equilibrium solution $n=-2$ at temperature
$\beta=6\beta_{c1}$. We plot both the equilibrium profile $\eta$ (dashed line)
and its unstable mode $\delta\eta$ (solid line) with eigenvalue $\lambda_1>0$.}
\label{figqVlasovm2}
\end{figure}

We display in fig.~\ref{figlambdabetaVlasov} the stability eigenvalues 
$\lambda_k$ for the equilibria $n=0$, $\pm 1$ and $\pm 2$. We can check
that the homogeneous solution $n=0$ shows the same stability properties as
those obtained in the thermodynamical and hydrodynamical approaches, as discussed
below eq.(\ref{Maa0}). On the other hand, from eq.(\ref{Maa0}) we can see that at
low temperatures we have for the highest eigenvalue $\tilde{\lambda}_1$ the 
asymptotic behaviour $\tilde{\lambda}_1 \simeq \sqrt{\beta/2\beta_{c1}}$ which 
yields $\lambda_1 \rightarrow \sqrt{2g\rhob}$. For the equilibrium $n=\pm 1$
we only plot the curves up to $3\beta_{c1}$ for clarity and because our formula 
(\ref{Res0}) for the residue contribution only holds close to $T_{c1}$. However,
we checked numerically that $\lambda_k$ remains negative at low $T$ (the
approximation (\ref{Res0}) only enters for $\lambda_k<0$ so that the sign of
the eigenvalues $\lambda_k$ does not depend on it). Therefore, as for the
thermodynamical and hydrodynamical approaches the states $\pm 1$ are stable
below $T_{c1}$.

\begin{figure}
\begin{center}
\epsfxsize=8 cm \epsfysize=6 cm {\epsfbox{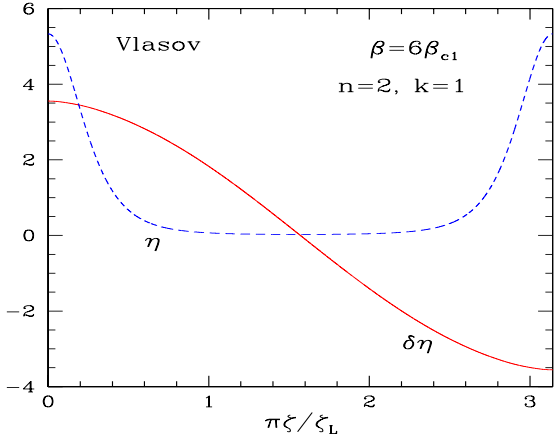}} 
\end{center}
\caption{The eigenmode $k=1$ for the equilibrium solution $n=2$ at temperature
$\beta=6\beta_{c1}$. We plot both the equilibrium profile $\eta$ (dashed line)
and its least stable mode $\delta\eta$ (solid line) with eigenvalue 
$\lambda_1 \simeq 0$.}
\label{figqVlasov2}
\end{figure}

In a similar fashion, the equilibrium $n=-2$ (which corresponds
to a density peak at $x=L/2$) shows one unstable mode as in 
fig.~\ref{figlambdabetaHydro} with $\lambda_1/\sqrt{2g\rhob}\rightarrow 1$ at
low temperature. However, fig.~\ref{figqVlasovm2} shows that the corresponding
eigenmode is significantly different from the one obtained in the hydrodynamical
framework (compare with fig.~\ref{figqHydrom2}). Indeed, the growing mode
$\delta\eta$ no longer tends to the derivative $\eta'$ associated with a pure
translation of the central density peak. Although close to the peak at $x=L/2$
it can be interpreted as a translation the perturbation extends as far as the
boundaries $x=0,L$, so that it also involves a deformation of the density
profile. Therefore the dynamics of the relaxation towards the stable equilibrium
$\pm 1$ proceeds in a manner which is different from the behaviour obtained
within the hydrodynamical model. In particular, the deformation implied by
this instability suggests that a solitonic behaviour cannot arise as some mass
leaks from the central peak down to the boundary $x=0$ while the peak moves 
towards $x=0$. This difference with respect to the hydrodynamical case may be
understood as follows. In the collisional fluid model the evolution proceeds in
a well-ordered manner through local processes, as each fluid element at location
$x$ reacts to slow changes in the local pressure and density. By contrast,
in the collisionless case described by the Vlasov equation particles probe far
away regions as their orbit extends to both sides of the density peak. Thus,
as the central peak moves towards $x=0$ as in fig.~\ref{figqVlasovm2} the
left/right symmetry is broken and the potential $\phi$ becomes deeper to the
left of the peak. Then, if we consider a subsample of particles which relax
to the new potential as $\rho_{\rm sub}\propto e^{-\phi}$, as they can move from
one side of the density peak to the other, they will populate the left side of
the peak more than the right side. This yields $\delta\eta(0)>0>\delta\eta(L)$
as in fig.~\ref{figqHydrom2} (of course the collisionless Vlasov dynamics 
conserves entropy but such a relaxation may be applied to the coarse-grained
distribution which mixes the two subsamples). 

Finally, the equilibrium $n=2$ also shows at first an unstable mode $\lambda_1$
which quickly goes towards zero as in fig.~\ref{figlambdabetaHydro} but contrary
to the hydrodynamical case the eigenvalue $\lambda_1$ becomes negative which
means that at low $T$ this state turns stable. The density profile of this
eigenmode displayed in fig.~\ref{figqVlasov2} at $6\beta_{c1}$ shows that it
is again significantly different from its hydrodynamical counterpart. Indeed,
as for equilibrium $-2$ it exhibits a broader extension than within the
hydrodynamical framework which is related to the collisionless nature of the
system which entails changes on a more global scale. This implies that 
$\eta+\delta\eta$ does not correspond to a quasi-static evolution where both 
density peaks would follow equilibrium profiles with different slowly varying 
masses. Thus, it appears that the decrease of entropy $S[f]$, understood as
an H-function as in Chavanis (2003), due to the distortion of the profiles
close to the boundaries $x=0,L$, farther from isolated equilibrium states
dominates the increase of entropy which would be associated with the 
transformation of the system from two separate density peaks towards only one
density peak (state $\pm 1$ which exhibits a higher $S[f]$, as seen from
fig.~\ref{figSE}). Thus, thermodynamical considerations or non-linear
analysis based on H-functions are not sufficient to obtain the dynamical
stability properties of the system. One needs to take into account the 
constraints brought by the Vlasov dynamics: one cannot select any arbitrary
deformation path from equilibrium $2$ towards equilibrium $\pm 1$. As discussed
in Lynden-Bell (1967), the infinite number of conserved quantities associated
with the collisionless dynamics (the total mass of phase-space element greater
than any $f$ is conserved) can actually be taken into account by using a
new ``collisionless entropy'' $\ln W$ which differs from the usual Boltzmann
entropy. On the other hand, the comparison
with sect.~\ref{Hydrodynamical-model} shows that the collisional and 
collisionless dynamics are significantly different, as it appears that
the hydrodynamical dynamics allows the system to flow from state $n=2$ to
state $n=\pm 1$ (which implies that equilibrium $n=2$ is unstable) whereas
at low $T$ the Vlasov dynamics prevents the system from going to $n=\pm 1$
from $n=2$ through linear instabilities (but clearly a non-linear perturbation
can lead to relaxation towards  $n=\pm 1$).

\section{Conclusion}
\label{Conclusion}

We have studied in this article the thermodynamical properties of a 1-D 
gravitational system built from the problem of large-scale structure formation
in cosmology. We have considered time-scales much smaller than the Hubble time
so that the expansion of the universe can be neglected and the system can be
restricted to a finite size $L$. Then, the cosmological background merely yields 
an effective external potential $V$ in addition to the gravitational 
self-interaction $\Phi$ so that an homogeneous equilibrium (corresponding to 
the Hubble flow) exists at all temperature $T$. We have found that all three 
micro-canonical, canonical and grand-canonical ensembles give the same results. 
There exists a series of critical temperatures $T_{cn}$ ($T_{c1}>T_{c2}>..$) 
below which two new equilibria $\pm n$ appear, reflecting the usual Jeans 
instability mechanism. Above $T_{c1}$ the homogeneous state is thermodynamically 
stable whereas below $T_{c1}$ it is unstable and only equilibria $\pm 1$ 
(corresponding to a density peak at the boundary $x=0$ or $L$) are stable
while other states $\pm n$ with $n\geq 2$ are unstable. The system shows a 
second-order phase transition at $T_{c1}$ where the specific heat is 
discontinuous.

Next, we have studied the hydrodynamical properties of these equilibria. We have
found that both the non-linear stability and the dynamical linear stability
coincide with the results of the thermodynamical analysis. The evolution of the
system can be understood in a simple manner from the translation and diffusion
of the various density peaks associated with states $\pm n$ which would relax
towards the stable equilibria $\pm 1$. Finally, we have studied the collisionless
dynamics governed by the Vlasov equation. The equilibria $\pm 1$ are again 
both linearly and non-linearly stable while the homogeneous state is unstable
below $T_{c1}$. We found that the equilibrium $n=-2$ (one density peak at 
$x=L/2$) is again unstable but the equilibrium $n=2$ (two density peaks located
at both boundaries) shows a new behaviour. Although it is unstable just below
$T_{c2}$ as in the thermodynamical and hydrodynamical approaches it now turns
stable at low temperatures. This departure from the results of the 
thermodynamical and hydrodynamical analysis is due to the dynamical constraints
associated with the equations of motion of a collisionless system. On the other 
hand, other equilibria $\pm n$ with $n\geq 3$ are again unstable. Therefore,
depending on the initial conditions the system would relax towards equilibria
$\pm 1$ or $2$ (i.e. a density peak at either one or both boundaries). 

Thus, the behaviour of the 1-D gravitational system (OSC) studied in this article
is quite similar to the HMF model, described for instance in 
Chavanis et al. (2005).
We obtain the same second-order phase transition from a high-temperature 
homogeneous equilibrium to a clustered equilibrium $\pm 1$ at low temperature.
The stability properties of these two configurations are also similar, for
all three thermodynamical, hydrodynamical and collisionless analysis. However,
in the case of the OSC system the scale-free nature of the
gravitational interaction leads to an additional series of equilibria $\pm n$
($n\geq 2$). These new states are unstable, except for the equilibrium $n=2$ 
which turns linearly stable for a collisionless dynamics at low $T$, but remains
unstable for both the thermodynamical and hydrodynamical approaches. This 
provides an interesting case where the collisionless dynamics exhibits a specific
behaviour and leads to a relaxation which clearly depends on initial conditions.
On the other hand, it would be interesting to study the non-linear evolution of 
the system and the possible merging of several density peaks in both the
hydrodynamical and collisionless cases. This might be studied through $N-$body
simulations as in Hohl \& Feix (1967) or through a direct computation of the 
Vlasov
dynamics as in Alard \& Colombi (2005). Finally, one should include the 
expansion of
the universe to study the non-equilibrium dynamics associated with cosmological
hierarchical scenarios. This is left for future studies but we can expect this
1-D gravitational system to provide a useful tool to investigate such processes.

\end{document}